\documentclass[12pt]{article}
\usepackage{epsf,rotating,latexsym,amssymb}           
\usepackage{mathptmx} 
\usepackage{times}
\usepackage{natbib}
\usepackage{lineno}
\newcommand{\mc}[3]{\multicolumn{#1}{#2}{#3}}

\newcommand{\QI}{\,Q_{\mathrm{I}}}
\newcommand{\QC}{\,Q_{\mathrm{C}}}

\linenumbers
\begin{document}
\bibliographystyle{astron}


\medskip

\begin{center}
{\Large \bf Three years of Ulysses dust data: 2005 to 2007}
\end{center}

\medskip

        H.~Kr\"uger$^{a,b,}$\footnote{{\em Correspondence to:} 
Harald Kr\"uger, krueger@mps.mpg.de}, 
        V.~Dikarev$^{c}$,
        B.~Anweiler$^b$,
        S.~F. Dermott$^d$,
        A.~L. Graps$^e$,
        E.~Gr\"un$^{b,f}$,
        B.~A.~Gus\-taf\-son$^d$,
        D.~P.~Hamilton$^g$, 
        M.~S.~Hanner$^h$,
        M.~Hor\'anyi$^f$,
        J.~Kissel$^a$,
        D.~Linkert$^{b}$,
        G.~Linkert$^b$,
        I.~Mann$^i$,
        J.~A.~M.~McDon\-nell$^j$,
        G.~E.~Mor\-fill$^k$, 
        C.~Polanskey$^l$,
        G.~Schwehm$^m$ and 
        R.~Srama$^{b,n}$

\medskip

\small
\begin{tabular}{ll}
a)& Max-Planck-Institut f\"ur Sonnensystemforschung, 37191 Katlenburg-Lindau, Germany \\
b)& Max-Planck-Institut f\"ur Kernphysik, 69029 Heidelberg, Germany\\
c)& Fakult\"at f\"ur Physik, Universit\"at Bielefeld, Postfach 100131, 33501 Bielefeld, Germany \\
d)& University of Florida, 211 SSRB, Campus, Gainesville, FL\,32609, USA \\
e)& Department of Space Studies, Southwest Research Institute, 1050 Walnut Street \\
  & Suite 300, Boulder, Colorado, 80302, USA \\
f)& Laboratory for Atmospheric and Space Physics, Univ. of Colorado, Boulder, \\ 
  & CO\,80309, USA\\
g)& University of Maryland, College Park, MD\,20742-2421, USA\\
h)& Astronomy Dept. 619 LGRT, University of Massachusetts, Amherst MA 01003, USA \\
i)& School of Science and Engineering, Kindai University, \\
  & Kowakae 3-4-1, Higashi-Osaka, Osaka, 577-8502, Japan \\
j)& Planetary and Space Science Research Institute, The Open University, \\ 
  & Milton Keynes, MK7 6AA, UK\\
k)& Max-Planck-Institut f\"ur Extraterrestrische Physik, 85748 Garching, Germany\\ 
l)& Jet Propulsion Laboratory, Pasadena, California 91109, USA\\
m)& ESAC, PO Box 78, 28691 Villanueva de la Ca{\~n}ada, Spain \\ 
n)& Universit{\"a}t Stuttgart, Institut f{\"u}r Raumfahrtsysteme, Pfaffenwaldring 31, \\ 
  & 70569 Stuttgart, Germany \\
\end{tabular}

\normalsize


\bigskip

\begin{abstract}

The Ulysses spacecraft has been orbiting the Sun on a highly inclined ellipse ($
i = 79^{\circ}$, perihelion distance 1.3~AU, aphelion distance 5.4~AU) since it
encountered Jupiter in February 1992. Since then it made almost three revolutions about the Sun. 
Here we report on the final three years of data taken by the on-board dust detector.
During this time, the dust detector recorded 609 dust impacts
of particles with masses
$10^{-16}\,\mathrm{g} \leq m \leq 10^{-7}\,\mathrm{g}$, bringing the mission total
to 6719 dust data sets. 
The impact rate varied from a low value of 0.3 per day at high ecliptic latitudes to
1.5 per day in the inner solar system. 
The impact direction of the 
majority of impacts between 2005 and 2007 is 
compatible with particles of interstellar origin, the rest are most 
likely interplanetary particles. 
We compare the interstellar dust measurements from 2005/2006 with the
data obtained during earlier periods (1993/1994) and (1999/2000) when Ulysses was traversing
the same spatial region at southern ecliptic latitudes but the solar cycle was at a 
different phase. During these three intervals the 
impact rate of interstellar grains varied by more than a factor of two. Furthermore, in 
the two earlier periods the grain impact direction was in agreement with the flow direction 
of the interstellar helium while in 2005/2006 we observed a 
shift in the approach direction of the grains by approximately $30^{\circ}$ away from the 
ecliptic plane. The reason for this shift remains unclear but may be connected with 
the configuration of the interplanetary magnetic field during solar maximum. 
We also find that  
the dust measurements are in agreement with the interplanetary flux model of 
\citet{staubach1997} which was developed to fit a 5-year span of Ulysses data.

\end{abstract}

\section{Introduction}

Ulysses was the only space mission so far that left the ecliptic plane
and flew over the poles of the Sun.
The spacecraft was launched in 1990 and was very successfully operated until 30 June 
2009, although individual instruments had to be turned off to conserve power.  
Its orbital plane was almost perpendicular to the ecliptic plane 
($79^{\circ}$ inclination) with 
an aphelion at Jupiter. This special orbit orientation allowed Ulysses to 
unambiguously detect
interstellar dust grains entering the heliosphere because the spacecraft's 
orbital plane was almost perpendicular to the flow
direction of the interstellar dust. Ulysses had a highly sensitive
impact ionisation dust detector on board which measured impacts of micrometre
and sub-micrometre dust grains. The detector was practically identical with 
the dust instrument which flew on board the Galileo spaceprobe. 
Both instruments were described in previous publications by
\citet{gruen1992a,gruen1992b,gruen1995a}. 

\subsection{Summary of results from the Ulysses dust investigations}

Comprehensive reviews of the scientific achievements of the 
Ulysses mission including results from the dust investigation 
were given by \citet{balogh2001} and \citet{gruen2001a}.
References to other works related to Ulysses and Galileo measurements on 
dust in the planetary system were also given by 
\citet{gruen1995c,gruen1995b,krueger1999a,krueger1999b,krueger2001a,
krueger2001b,krueger2006a,krueger2006b,krueger2009a}. Some mission highlights
are also summarized in Table~\ref{papers}.

Various dust populations were investigated with the Ulysses and Galileo dust 
experiments in interplanetary space: the interplanetary dust complex including
$\beta$-meteoroids (i.e. dust particles which leave the solar system on 
unbound orbits due to acceleration by radiation pressure), interstellar grains 
sweeping through the heliosphere, and dust stream particles expelled from the 
jovian system by electromagnetic forces, to name only the most significant dust 
types detected with Ulysses which have been analysed 
so far. In the following, we summarise the most significant achievements of the
Ulysses dust measurements. 

Ulysses and Galileo dust measurements were used to study the 
3-dimensional structure of the interplanetary dust complex 
and its relation to the underlying populations of parent bodies like 
asteroids and comets \citep{divine1993,gruen1997a,staubach1997}. Studies of 
asteroidal dust released from the IRAS dust bands show that 
they are not efficient enough dust sources to maintain the observed
interplanetary dust cloud \citep{mann1996b}. The 
state of the inner solar system dust cloud within approximately 
1~AU from the Sun including dust destruction and ion formation 
processes in relation to so-called solar wind
pickup ions detected by the Solar Wind Ion Composition 
Spectrometer (SWICS) onboard Ulysses was investigated 
\citep{mann2004,mann2005}. An improved 
physical model was developed for  the interplanetary meteoroid 
environment \citep{dikarev2001b,dikarev2002b,dikarev2005} which uses 
long-term particle dynamics to define individual 
interplanetary dust populations. The Ulysses and Galileo in-situ 
dust data are an important data set for the validation of this model.
The properties of $\beta$-meteoroids 
 were also studied with the 
Ulysses data set \citep{wehry1999,wehry2004}. Finally, the potential
connection of cometary dust trails and enhancements of the
interplanetary magnetic field as measured by Ulysses was discussed 
by \citet{jones2003b}.

During its first flyby at Jupiter in 1992, the Ulysses dust instrument 
discovered burst-like intermittent streams 
of tiny dust grains in interplanetary space \citep{gruen1993a} 
which had been emitted from the jovian system \citep{hamilton1993a,horanyi1993a,zook1996}. 
This discovery was completely unexpected as no periodic phenomenon 
for tiny dust grains in interplanetary space was previously known. These 
grains strongly interacted with the interplanetary and the jovian magnetic 
fields \citep{horanyi1997,gruen1998} 
and the majority of them originated from Jupiter's moon Io \citep{graps2000a}.
In February 2004 Ulysses had its second Jupiter flyby (at 0.8~AU
distance from the planet) and again measured the jovian dust streams
\citep{krueger2005b,krueger2006c,flandes2007,flandes2009}.

Another important discovery made with  Ulysses  were interstellar 
dust particles sweeping through the heliosphere \citep{gruen1993a}. 
The grains which originated from the very local interstellar environment of
our solar system  
were identified by their impact direction and impact 
velocities, the latter being compatible with particles moving 
on hyperbolic heliocentric trajectories \citep{gruen1994a}. Their
dynamics depends on the grain size and is strongly affected by the 
interaction with the interplanetary magnetic field and by solar 
radiation pressure 
\citep{landgraf1999a,landgraf2000b,mann2000a,czechowski2003a,czechowski2003b,landgraf2003}. 
As a result, the size distribution and fluxes of grains measured inside the 
heliosphere are strongly modified.
Studies of the dust impacts detected with both Ulysses and Galileo showed that
the intrinsic size distribution of interstellar grains in the local interstellar
environment of our solar system 
extends to grain sizes larger than those detectable by astronomical 
observations \citep{frisch1999a,frisch2003a,landgraf2000a,gruen2000b}. 
Observations of radar meteors entering the Earth's atmosphere at high speeds also 
indicate the existence of even larger interstellar
grains \citep{taylor1996b,baggaley2002}. 

The Ulysses and Galileo interstellar dust measurements showed that the 
dust-to-gas mass ratio in the local interstellar cloud is 
higher than the standard interstellar value derived from 
cosmic abundances \citep{landgraf1998a,frisch1999a}. This implied the existence of inhomogeneities 
in the diffuse interstellar medium on relatively small length scales.
In 2005/2006 the Ulysses measurements showed a $30^{\circ}$ 
shift in the impact direction of interstellar grains with 
respect to the interstellar helium flow 
\citep{krueger2007b}. The reason of this shift is presently unclear. 
Finally, it turned out that the dust sensor 
side walls have a similar sensitivity to dust impacts
as the detector target itself \citep{altobelli2004a,willis2005}. This shows that 
earlier investigations which neglected the contributions of the side wall
overestimated the interstellar dust flux by about 20\%. Since the contribution
of the sensor side wall increases the effective dust instrument field-of-view, the 
velocity dispersion of the observed interstellar dust stream turned out 
to be smaller by about 30\% than previously thought \citep{altobelli2004a}.  

Due to its unique highly inclined heliocentric trajectory Ulysses was able to monitor 
interstellar dust at high ecliptic 
latitudes between 3 and 5~AU. Dust measurements between 0.3 
and 3~AU in the ecliptic plane exist also from Helios, Galileo 
and Cassini. This data shows evidence for distance-dependent 
alteration of the interstellar dust stream caused by radiation 
pressure, gravitional focussing and electromagnetic interaction 
of the grains with the time-varying interplanetary magnetic field 
\citep{altobelli2003,altobelli2005a,altobelli2005b}.

\subsection{The Ulysses and Galileo dust data papers}

The Ulysses dust detector obtained dust data for more than 17 years, making
it the longest continually-operating spacebourne dust detector to date. 
With the publication of this paper, the full Ulysses data set, along
with 13 years of Galileo data, will be fully described in the 
scientific literature.
The reduction process of Ulysses and Galileo dust data was described by 
\citet[][hereafter Paper~I]{gruen1995a}.
In the odd-numbered Papers~III, V, VII and IX 
\citep{gruen1995c,krueger1999b,krueger2001b,krueger2006b} we 
presented the Ulysses data set spanning the time period from launch in 
October 1990 to December 2004. The companion even-numbered Papers~II, IV, VI and VIII
\citep{gruen1995b,krueger1999a,krueger2001a,krueger2006a}
discussed the ten years of Galileo data from October 1989 to December 
1999. The current paper (Paper~XI) extends the Ulysses data set 
from January 2005 until the Ulysses dust measurements ceased in 
November 2007, and a companion paper 
\citep[][Paper~X]{krueger2010a} presents Galileo's final 
measurements at Jupiter from 2000 to 2003. A summary of the temporal and spatial
coverage of our Ulysses data papers is given in Table~\ref{papers}.

The main data products are a table of the impact 
rate of all impacts determined from the particle accumulators and a table 
of both raw and reduced data of all dust impacts for which the 
full data set of measured impact parameters was transmitted to Earth.
The information presented in these papers is similar to data which we 
have submitted to the various data archiving centres (Planetary 
Data System, NSSDC, Ulysses Data Centre). Electronic access to the 
data is also possible via the world wide web: 
http://www.mpi-hd.mpg.de/dustgroup/.

This paper is organised like our earlier Papers~III, V, VII and IX. We begin with an 
overview of important events of the Ulysses mission between 2005 and 2007 
(Section~\ref{mission}). Sections~\ref{events} and \ref{analysis} describe
and analyse the Ulysses dust data set for this period. 
In Section~\ref{discussion} we discuss the dust data set from the entire 
Ulysses mission, we analyse in particular the interstellar dust measurements 
obtained in the 2005 to 2007 interval, and we compare these results to 
previous ones. In Section~\ref{summary} we summarise our conclusions. 

\section{Mission and instrument operation} \label{mission}

\subsection{Ulysses mission and dust instrument characteristics}

\begin{figure}[h]

\vspace{-16cm}

\parbox{0.99\hsize}{
\hspace{-3cm}
\includegraphics[scale=0.95]{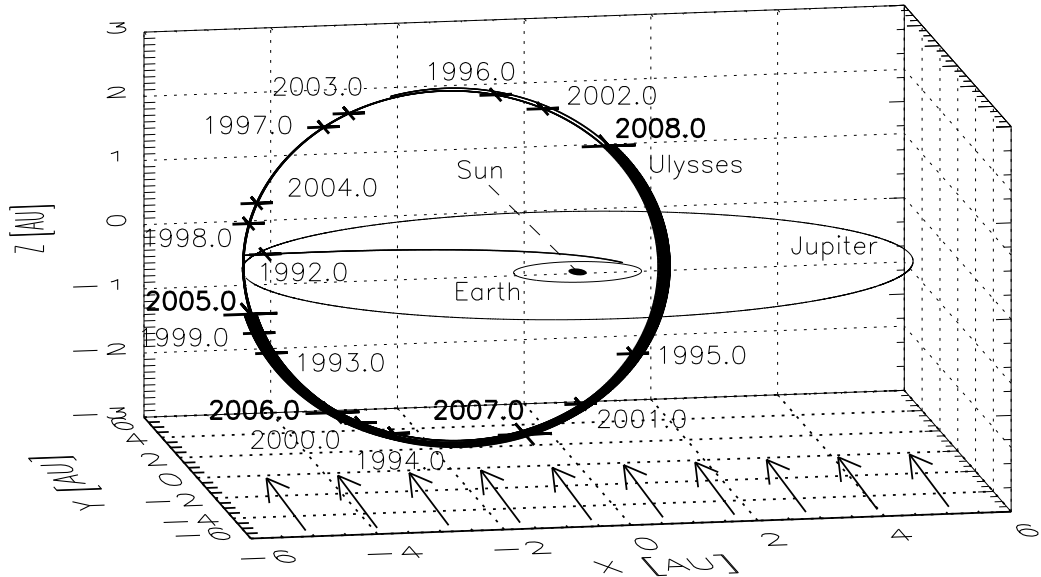}
}

        \caption{\label{trajectory}
The trajectory of Ulysses in ecliptic coordinates with the Sun at the centre. 
The orbits of Earth and Jupiter indicate the ecliptic plane, and the initial
trajectory of Ulysses was in this plane. Since Jupiter flyby 
in early 1992 the orbit has been almost perpendicular to the ecliptic plane 
(79$^{\circ}$ inclination).
Crosses mark the spacecraft position at the beginning 
of each year. The 2005 to 2007 part of the trajectory is shown as a 
thick line. Vernal equinox is to the right (positive x axis). Arrows indicate the
undisturbed interstellar dust flow direction which is within the measurement 
accuracy co-aligned with the direction of the interstellar helium gas flow. 
It is almost perpendicular to the orbital plane of Ulysses.
}
\end{figure}

The Ulysses spacecraft was launched on 6 October 1990. 
A swing-by manoeuvre at Jupiter in February 1992 rotated the orbital 
plane 79$^{\circ}$ relative to the ecliptic plane. On the resulting 
trajectory (Figure~\ref{trajectory}) Ulysses finished two full 
revolutions about the Sun. Passages over the south pole 
of the Sun occurred in October 1994, November 2000 and February 2007, passages
through the ecliptic plane at a perihelion distance of 1.3~AU 
occurred in March 1995, May 2001 and August 2007, and passes over the 
Sun's north pole were in August 1995, October 2001 and January 2008. In April 1998 
and July 2004 the spacecraft crossed the ecliptic plane at an 
aphelion distance of 5.4~AU. Orbital elements for the out-of-ecliptic part of the 
Ulysses trajectory are given in Paper~VII.
Figure~\ref{trajectory} shows that during most of the time interval 
considered in this paper Ulysses was at southern ecliptic latitudes.

\begin{figure}
\vspace{-14cm}

\parbox{0.99\hsize}{
\hspace{-3cm}
\includegraphics[scale=0.9]{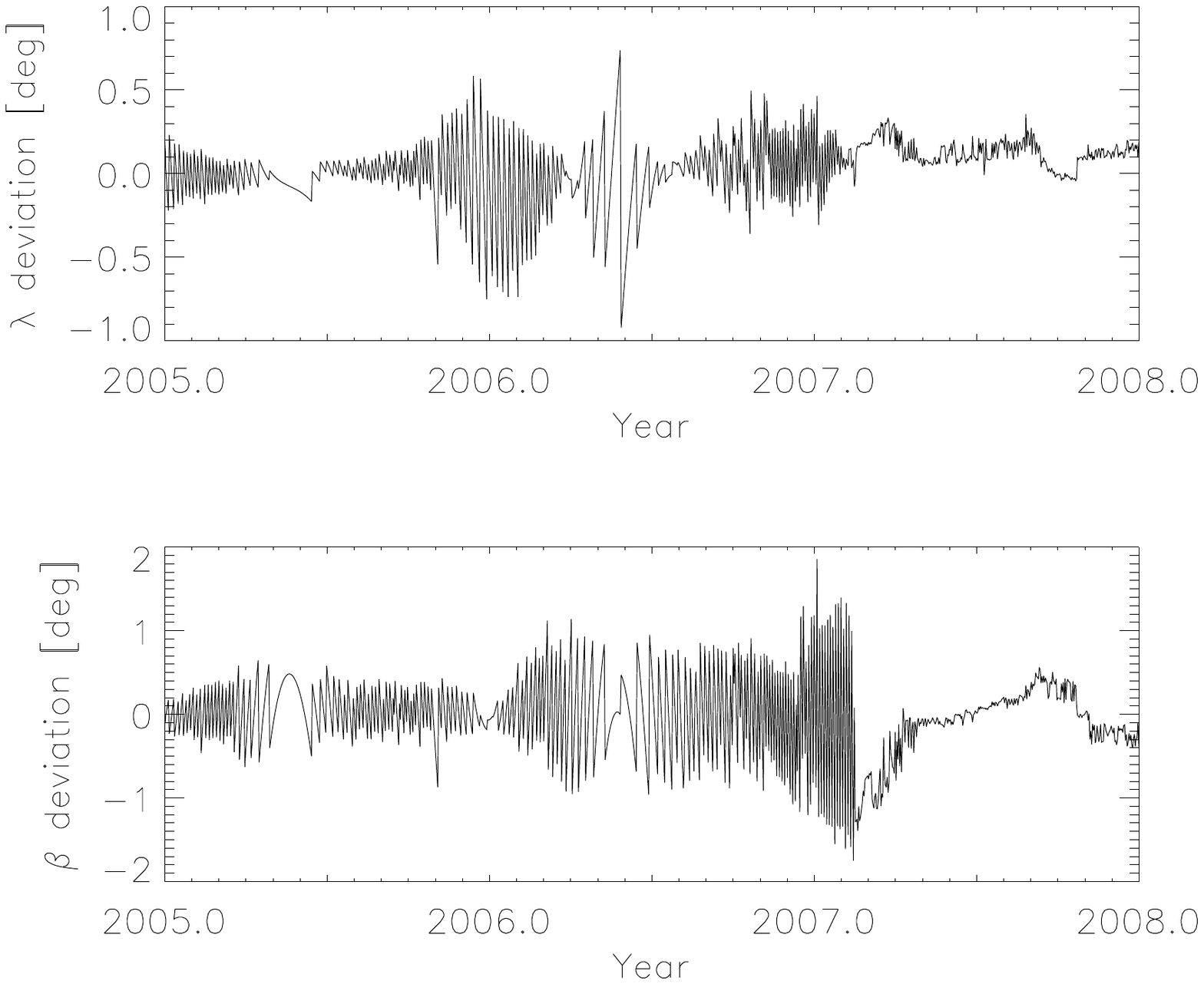}
}
        \caption{\label{pointing}
Spacecraft attitude: deviation of the antenna pointing direction 
(i.e. positive spin axis) from the nominal Earth direction. The angles are 
given in ecliptic longitude (top) and latitude (bottom, equinox 1950.0).
}
\end{figure}

Ulysses is a spinning spacecraft, and the dust sensor orientation 
at the time of a 
dust particle impact is recorded, allowing for an independent determination
of the grain impact direction. 
Ulysses spins at five revolutions per minute about the centre line 
of its high gain antenna which normally points at Earth. 
Figure~\ref{pointing} shows the deviation of the spin axis 
from the Earth direction for the period 2005 to 2007.
Most of the time the spin axis pointing was within 
$1^{\circ}$ of the nominal Earth direction, similar to
the mission before 2005. This small deviation 
is usually negligible for the analysis of measurements with the 
dust detector.
The Ulysses spacecraft and mission are explained in more detail by 
\citet{wenzel1992}. Details about the data transmission to Earth 
can also be found in Paper~III.

The Ulysses dust detector (GRU) has a 140$^{\circ}$ wide field of view and
is mounted at the spacecraft nearly at right angles (85$^{\circ}$) to the antenna 
axis (spacecraft spin axis). Due to this mounting geometry, the dust sensor 
is most sensitive to particles approaching from the plane perpendicular to the 
spacecraft-Earth direction. The impact direction of dust particles is 
measured by the rotation angle which is the sensor viewing direction at the 
time of a dust impact. During one spin revolution of the spacecraft the 
rotation angle scans through a complete circle of 360$^{\circ}$. 
Zero degrees rotation angle is defined to be the 
direction closest to ecliptic north. At high ecliptic latitudes, however, the 
sensor pointing at 0$^{\circ}$ rotation angle significantly deviates from the 
actual north direction. During the passages over the Sun's polar regions the sensor 
always scans through a plane tilted by about $30^{\circ}$ from the ecliptic 
plane and all rotation angles lie close to the ecliptic plane 
\citep[cf. Figure~4 in][]{gruen1997a}. A sketch of the viewing 
geometry around aphelion passage can be found in 
\citet{gruen1993a}.

The available electrical power on board Ulysses became an issue beginning 
in 2001 due to decreasing power generation of the radioisotope batteries 
(RTGs). Some instrument heaters on board had to be switched off to save power 
which in turn also reduced the on board temperature. Later, in 2002, power 
consumption could not be sufficiently reduced anymore by simply 
switching off heaters, and a cycling instrument operation scheme had to be 
implemented: one or more of the scientific instruments had to be switched off
at a time. In the 2005 to 2007 interval considered in this paper the dust instrument was 
switched off during a total time period of about seven months, separated into three
individual time periods (see Sect.~\ref{sec_instr_operation}).
After 30 November 2007 the dust instrument remained switched off 
permanently. It was planned to switch the dust instrument
on in January 2008 again, but due to a failure on 
board the spacecraft, this did not happen. Hence, no dust data were
obtained after 30 November 2007 although operation of the Ulysses spacecraft 
continued until 30 June 2009.
 
\subsection{Dust instrument operation}

\label{sec_instr_operation}

Table~\ref{event_table} gives significant mission and dust instrument 
events from 2005 to 2007. Earlier events are only listed if especially significant.
A comprehensive list of events from launch until the end of 2004 was given in 
Papers~III, V, VII and IX. 

During the earlier Ulysses mission phases until 2000, several spacecraft anomalies 
occurred during which all scientific instruments on board were 
switched off automatically (Disconnection of all Non-Essential Loads 
-- DNELs for short). No such anomaly occurred in the 2005 to 2007 interval.
The dust instrument was switched off 
between 28 September 2006 and 9 March 2007, from 2 April 2007 to 30 April 2007
and after 30 November 2007 ({\em cf.} Table~\ref{papers}), respectively, because there 
was not enough electical power generated by the RTGs anymore to operate all 
instruments simultaneously.

The dust instrument has two heaters to allow for a relatively 
stable operating temperature within the sensor. By heating one of 
the two or both heaters, three different heating power levels can 
be achieved (0.4~W, 0.8~W or 1.2~W; for comparison, the total
power consumption of the instrument without heaters is 2.2~W). Sensor 
heating was necessary when the spacecraft was outside about 2~AU 
because relatively little radiation was received from the Sun. 

Before 2001 one or both heaters were switched on beyond 2~AU,
except close to the Sun when both were switched off.
Since November 2001 the maximum allowed heating power
for nominal dust instrument operation has been limited to 0.8~W 
to save power on board Ulysses. This reduced 
the sensor temperature by about $\rm 10^{\circ}\,C$ 
as compared to the configuration with 1.2~W heating power at 
similar heliocentric distance. The full heating power was only 
allowed for short periods before instrument switch on to avoid 
damage of the electronics. In 2005 and 2006 the instrument was 
operated with 0.8~W heating power, and in 2007 the heating power
was set to 0.4~W, except during switch-ons when it was raised to 1.2~W.

Table ~\ref{event_table} lists the total powers consumed
by the heaters. 
From 2005 to 2007 the temperature of the dust sensor was between 
$-37^{\circ}\,\rm C$ and $+16^{\circ}\,\rm C$. The  lower limit
for the specified operational range is 
$\rm - 30^{\circ}\,\rm C$. No major effect of the reduced temperature
was recognised (for an anomalous temperature-related flipping in the housekeeping value of 
the channeltron high voltage in 2004 see Paper IX).  

No reprogramming of the Ulysses dust instrument occurred in the 2005 to 2007
interval. In the earlier mission the instrument was reprogrammed three times
and the reader is referred to Papers  V and IX for details.
In particular, a new classification 
scheme for impact events was implemented in April 2002 which is the same
as the one installed in the Galileo instrument in July 1994 (Paper~IV).

\subsection{Instrument sensitivity and noise}    \label{noise}

Analysis of the in-orbit noise characteristics of the dust
instrument (Paper~III) led to a relatively noise-free configuration with which 
the instrument was normally operated until August 2000: 
channeltron voltage 1140~V (HV~=~3); 
event definition status such that either the 
channeltron or the ion-collector channel could, independent of each other,
start a measurement cycle 
(EVD~=~C,~I); detection thresholds for ion-collector, channeltron and 
electron-channel set to the lowest levels and the detection threshold for 
the entrance grid set to the first digital step (SSEN~=~0,~0,~0,~1). See 
Paper~I for a description of these terms. 

During the entire Ulysses mission dedicated noise tests were performed at monthly intervals
in order to monitor instrument health and noise characteristics.
During all these tests the 
operational settings were changed in four steps at one-hour intervals, 
starting from the nominal configuration described above: 
a) set the event definition status such that the channelton, the ion 
collector and the electron-channel can initiate a measurement cycle 
(EVD~=~C,~I,~E);
b) set the thresholds for all channels to their lowest levels 
(SSEN~=~0,~0,~0,~0);
c) reset the event definition status to its nominal configuration (EVD~=~C,~I) 
and increase the channeltron high voltage by one step with respect 
to the nominal configuration; 
d) reset the instrument to its nominal configuration (i.e. reduce the
channeltron high voltage by one step and set the detection thresholds to 
SSEN~=~0,~0,~0,~1).
 
The noise tests performed in the earlier mission before 2000 revealed a long-term drop in the 
noise sensitivity of the instrument which was most likely caused by a reduction 
in the channeltron amplification 
due to electronics degradation \citep[Paper~IX and][]{krueger2005a}.
To counterbalance the reduced amplification
we increased the channeltron high voltage by one digital step 
in August 2000 (HV = 4, 1250~V) which raised the 
instrument sensitivity close to its original value from the earlier mission again
(Paper~IX).
Although the aging of the channeltron led to a drop in the number of 
class~3 impacts, the dust 
impacts which have caused these events should have shown up 
in lower quality classes.
The noise response of the Ulysses dust detector was monitored since then
and no significant drop in the sensitivity was recognized. In the time
interval considered here HV=4, EVD=C,I, SSEN=0001 was the nominal configuration
of the dust instrument 
(during noise tests the channeltron voltage was always raised by 
one digital step, i.e. to HV = 5, 1370~V).

\begin{figure}
\vspace{-2cm}

\hspace{-1cm}
\parbox{0.99\hsize}{
\parbox{0.85\hsize}{
\includegraphics[scale=0.7]{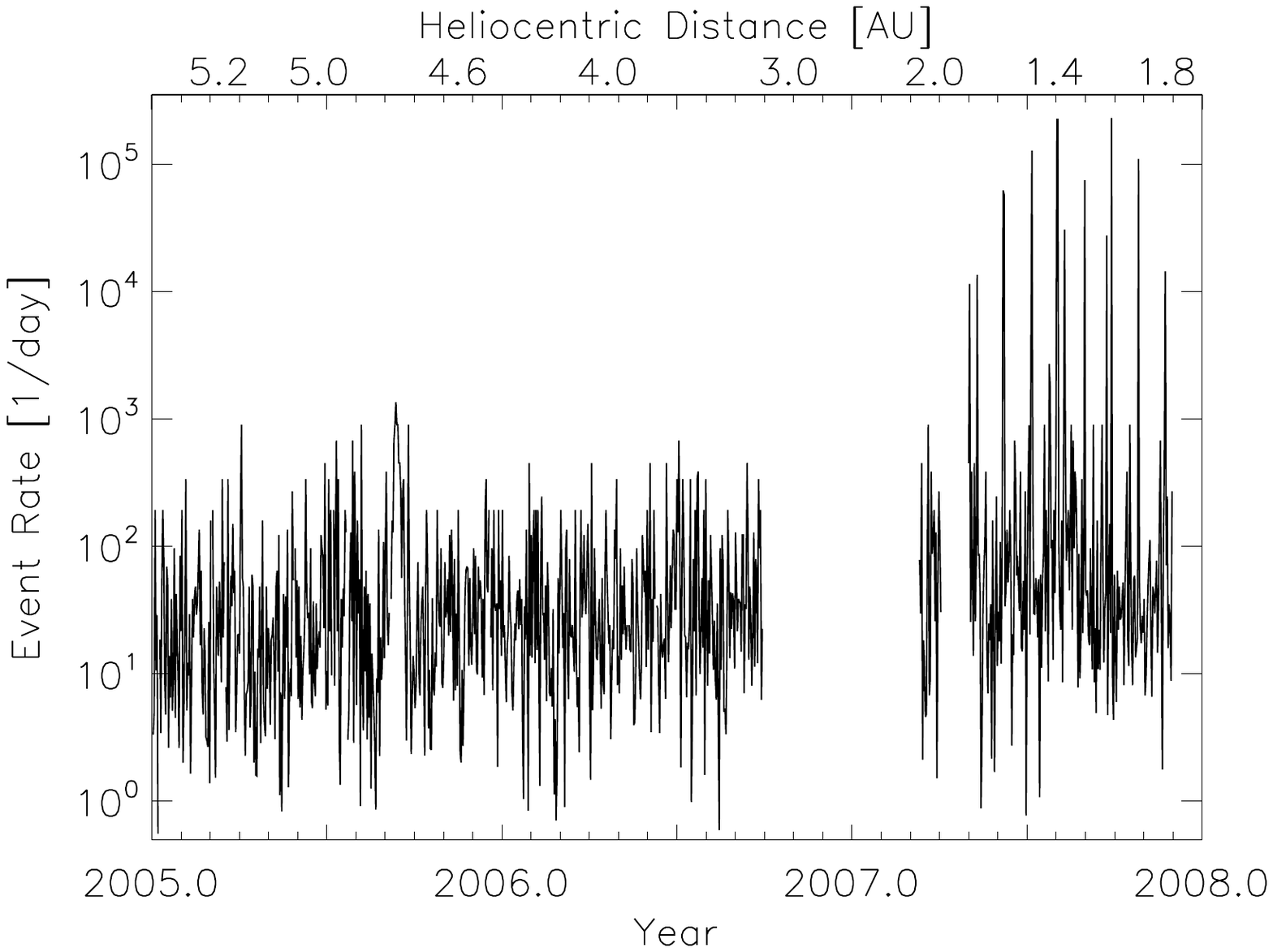}
}

\vspace{-14.5cm}
\parbox{0.85\hsize}{
\includegraphics[scale=0.7]{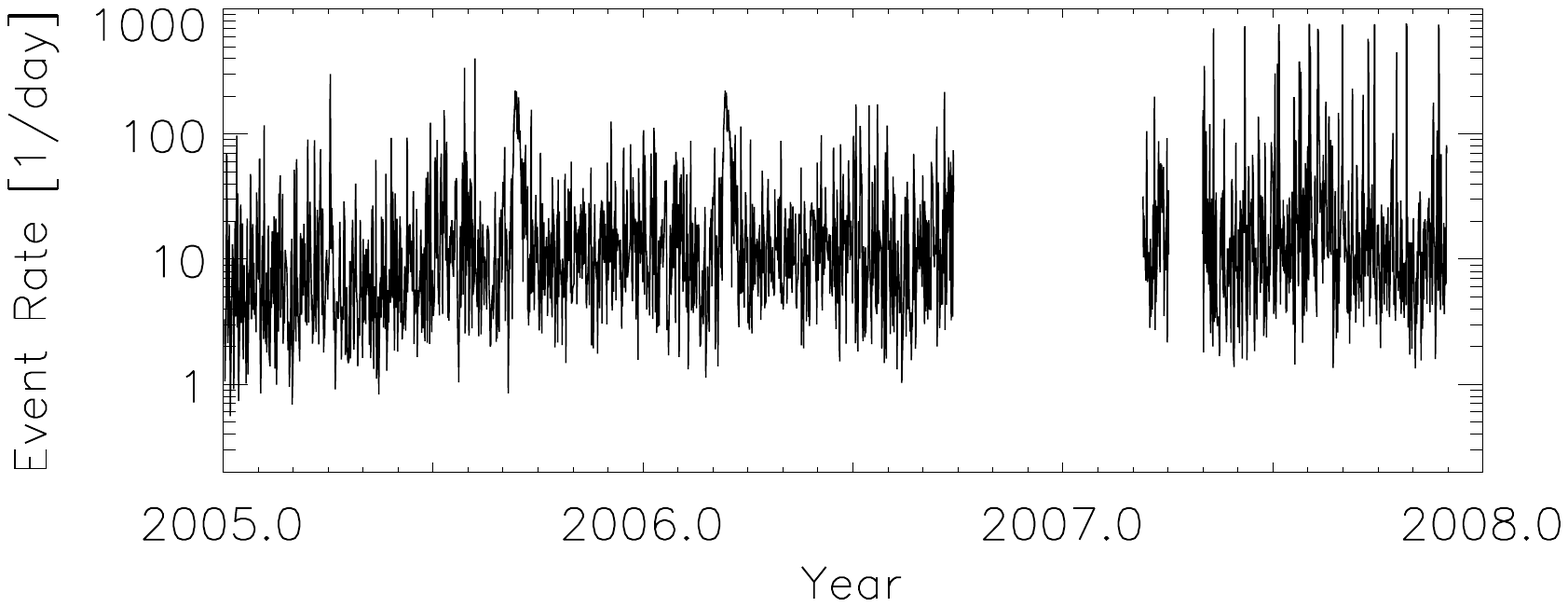}
}
\vspace{-9cm}
}
        \caption{\label{noiserate}
Noise rate (class~0 events) detected with the dust instrument. 
The heliocentric distance of Ulysses is indicated at the top.
Upper panel: Daily 
maxima in the noise rate (determined from the AC01 accumulator). 
Sharp spikes which are strongest after May 2008 when Ulysses was
in the inner solar system are caused by periodic noise tests.
Lower panel: One-day average of the noise rate
calculated from the number of AC01 events for which the complete 
information was transmitted to Earth.
}
\end{figure}

Figure~\ref{noiserate} shows the noise rate of the dust instrument 
for the 2005 to 2007 period. 
The upper panel shows the daily maxima of the noise rate. In the earlier mission 
before November 2001 the daily maxima were dominated 
by noise due to interference with the sounder of the Unified RAdio and 
Plasma wave instrument (URAP) on board 
Ulysses \citep{stone1992}. 
Since November 2001 the sounder was switched off permanently for power
saving and no sounder noise occurred anymore (see Papers~III, V, VII and
IX for details of the sounder operation and sounder-related noise).
Individual sharp spikes in the upper panel of Figure~\ref{noiserate}
are caused by noise tests which occurred at approximately monthly intervals,
They are best seen in 2007 when Ulysses was close to the Sun which is known to 
be a strong source of noise. 

The bottom panel of Figure~\ref{noiserate} shows the daily averages in the
noise rate. The average was about 10 events 
per day at random times and it shows that dead time is 
negligible. These noise rates are very similar
to those measured since August 2000 (Paper~IX) when the instrument was operated with
the same channeltron voltage as in the 2005-07 interval, implying
that no significant channeltron degradation has occurred since 2000. 

\section{Impact events} \label{events}

The dust instrument classifies all impact events into four classes and six 
ion charge amplitude 
ranges which leads to 24 individual categories, with one accumulator belonging 
to one individual 
category. Class~3, our
highest class, are real dust impacts and class~0 are mostly noise events. 
Depending upon the noise of the charge measurements, classes~1 and 2 
can be true dust impacts or noise events. Two classification schemes were used
in the Ulysses dust instrument: 
since 26 March 2002, when the dust instrument was reprogrammed,  
the classification scheme has been 
the same as the one used in the Galileo instrument (described in Paper~IV).
A different scheme was used before which is described in Paper~I.  

Between 1 January 2005 and 30 November 2007 the complete data sets (sensor 
orientation, charge amplitudes, charge rise times, etc.) of 6970
events 
including 609 dust impacts were transmitted to Earth. Table~3 
lists the number of all dust impacts counted with the 24 accumulators of 
the instrument. `AC{\em xy}' refers to class number `$x$' and amplitude
range `$y$' (for a detailed description of the accumulator categories
see Paper~I). As discussed in the previous section, most noise events were 
recorded during the time 
periods when the dust instrument was configured to its high sensitive state 
for noise tests. During these periods many events were only counted 
by one of the 24 accumulators
because their full information was overwritten before the data could be
transmitted to Earth. Since the dust impact
rate was low during times outside these periods, it is expected
that only the data sets of very few true dust impacts were lost. 


Table~4 lists all 609 particles detected 
between January 2005 and November 2007 for which the complete information exists.
Note that approximately 45 jovian stream particles were detected in six dust streams
in 2005 \citep[AR1,][]{krueger2006c}
which are about 10~nm in size and their velocities exceed 
$\rm 200\,km\,s^{-1}$ \citep{zook1996}.  
Their mass and speed calibration is unreliable because
their masses and speeds are outside of the calibration range of the dust
instrument. 

In Table~4 dust particles are identified by their sequence number 
and their impact time. 
The event category -- class (CLN) 
and amplitude range (AR) -- are given. Raw data as transmitted to Earth 
are displayed in the next columns: sector value (SEC) which is the
spacecraft spin orientation at the time of impact, 
impact charge numbers (IA, EA, CA) and rise times (IT, ET), time
difference and coincidence of electron and ion signals (EIT, EIC),
coincidence of ion and channeltron signal (IIC), charge reading at
the entrance grid (PA) and time (PET) between this signal and
the impact. Then the instrument configuration is given: event
definition (EVD), charge sensing thresholds (ICP, ECP, CCP, PCP) and
channeltron high voltage step (HV). See Paper~I for further
explanation of the instrument parameters. 

The next four columns in Table~4 give information about Ulysses' orbit: 
heliocentric distance (R), ecliptic longitude and latitude (LON, LAT) 
and distance from Jupiter ($\rm D_{Jup}$, in astronomical units).
The next column gives the 
rotation angle (ROT) as described in Section~\ref{mission}. 
Then follows the pointing direction of the dust instrument at 
the time of particle impact in ecliptic longitude and latitude 
($\rm S_{LON}$, $\rm S_{LAT}$).
Mean impact velocity ($v$, in $\rm km\,s^{-1}$) and velocity error factor 
(VEF, i.e. multiply or
divide stated velocity by VEF to obtain upper or lower limits) as well 
as mean particle mass ($m$, in grams) and mass error factor (MEF) are given in the 
last columns. For VEF $> 6$, both velocity and mass values should be
discarded. This occurs for 72 impacts. No intrinsic dust charge 
values are given \citep[see][for a detailed analysis]{svestka1996}.
Recently, reliable charge measurements for interplanetary dust grains 
were reported for the Cassini dust detector \citep{kempf2004}. 
These measurements may lead to an improved understanding of the charge 
measurements of Ulysses and Galileo in the future. 

\bigskip

\section{Analysis} \label{analysis}

\begin{figure}
\vspace{-6cm}
\parbox{0.99\hsize}{
\hspace{-2cm}
\includegraphics[scale=0.8]{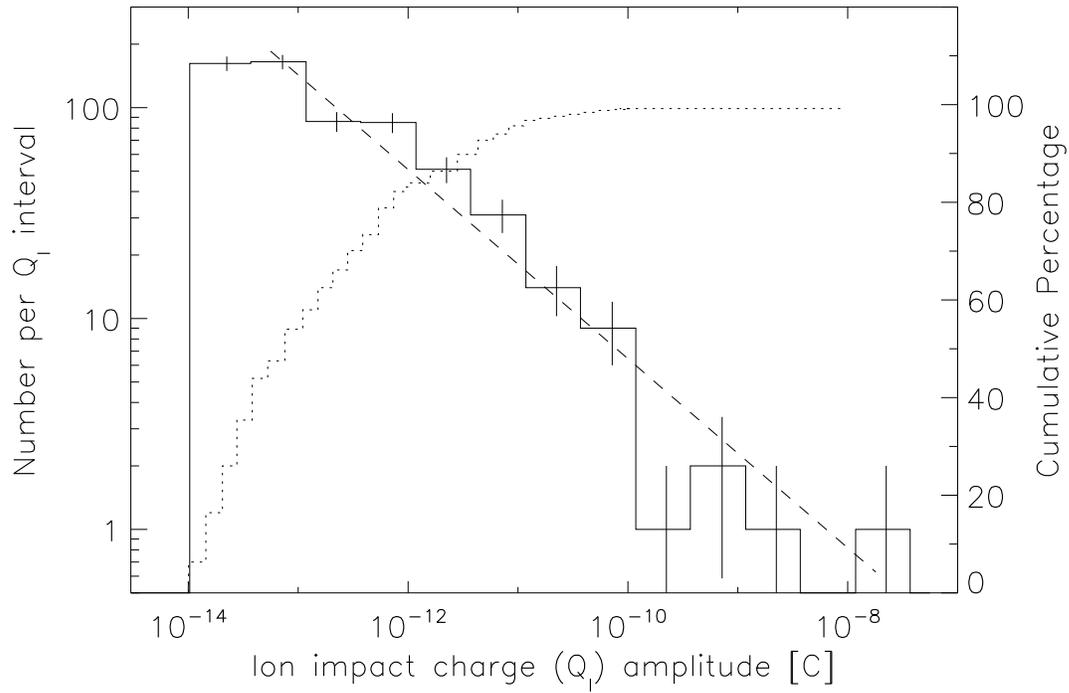}
}
\vspace{-9cm}
        \caption{\label{nqi}
Distribution of the impact charge amplitude $\QI$ for all 
dust particles 
detected from 2005 to 2007. The solid line
indicates the number of impacts per charge interval, and the 
dotted line shows the cumulative percentage. 
Vertical bars
indicate the $\sqrt{n}$ statistical fluctuation.  A 
power law fit to the data with $\QI > 3 \cdot 10^{-14}\,\mathrm{C}$ is
shown as a dashed line (Number $N \sim \QI^{-0.45}$). 
}
\end{figure}

The most important impact parameter determined by the dust instrument is
the positive charge measured on the ion collector, $\QI$, 
because it is relatively insensitive to noise. Figure~\ref{nqi} shows the 
distribution of $\QI$ for all dust particles detected from 2005 to 
2007. Ion impact charges have been detected over the entire range of six 
orders of magnitude in impact charge that the dust instrument can measure. 
One impact is close to the saturation limit of $\rm \sim 10^{-8}\,C$
and may thus constitute a lower limit of the actual impact charge.
The impact charge distribution of the big particles ($ \QI > 10^{-13}\,\mathrm{C}$)
follows a power law distribution with index $-0.45$ and is shown as a dashed
line. This value is very similar to  earlier Ulysses measurements of 
2000 to 2004 (-0.40; cf. Paper~IX).

In the earlier 1993 to 2004 data set (Papers~V, VII and IX) the impact charge 
distribution was reminiscent of three individual populations: 
small particles with impact charges $ \QI \rm <10^{-13}\,C$ (AR1), intermediate
size particles with $ 10^{-13}\,\mathrm{C} \leq  \QI  \leq  10^{-11}\,\mathrm{C}$ 
(AR2 and AR3) and big particles with $\QI \rm >  10^{-11}\,C$ 
(AR4 to AR6). This is also visible in the present data set, although 
less pronounced. The intermediate 
particles are mostly of interstellar origin and the big particles are 
attributed to interplanetary grains 
\citep[][see also Section~\ref{discussion}]{gruen1997a}. The small particle 
impacts (AR1) detected over the polar 
regions of the Sun are candidates for being 
interplanetary $\beta$-meteoroids \citep{hamilton1996a,wehry1999,wehry2004}.

It should be noted that the charge distribution shown in Figure~\ref{nqi}
is very similar to the one measured with Galileo in interplanetary space
between 1993 and 1995 (i.e. between 1 and 5 AU; Paper~IV). In particular, 
the power law index of $-0.43$ for the big particles was practically 
identical. This indicates that both dust instruments basically 
detected the same dust populations in interplanetary space and that their
responses to dust impacts are very similar.
The only significant difference is a dip in 
the Ulysses charge distribution at $\rm 2 \cdot 10^{-10}\,C$
(Figure~\ref{nqi}) which is also evident in all earlier Ulysses data
(Papers~III, V, VII and IX). In particular, it is visible in both the
ecliptic and out-of-ecliptic phases of Ulysses which indicates that it 
is not sensitive to the impact speed or the dust population of the grains.
A small but much weaker dip is also seen in some but not all Galileo data sets.
We therefore conclude that the dip is most likely
due to an artefact in the Ulysses instrument electronics in AR5.

\begin{figure}
\vspace{-6cm}
\parbox{0.99\hsize}{
\hspace{-1cm}
\includegraphics[scale=0.8]{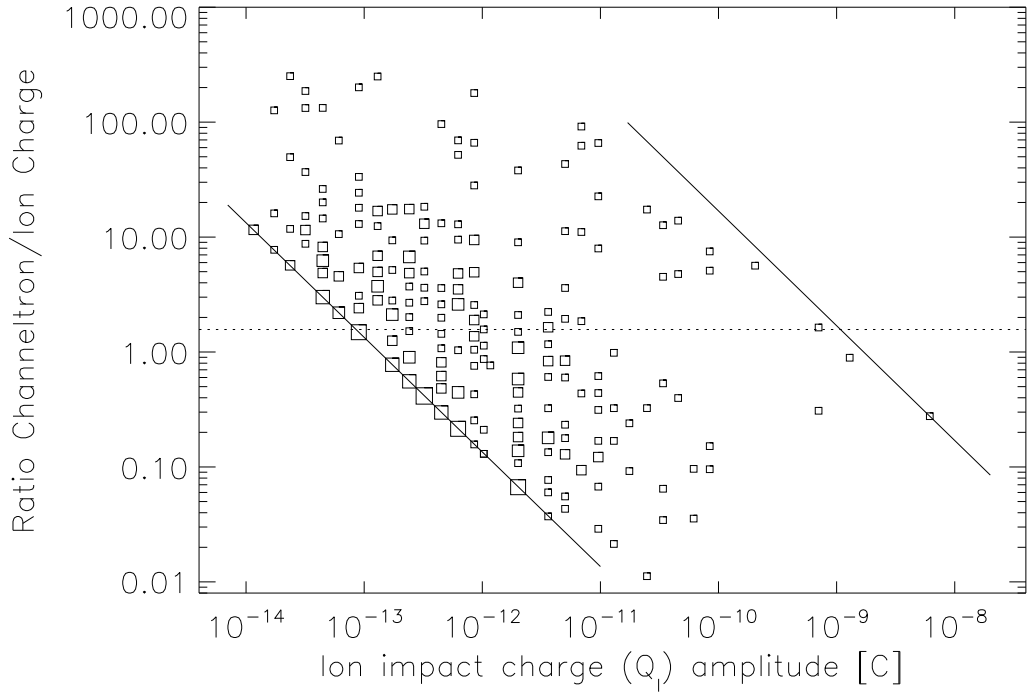}
}
\vspace{-9cm}
        \caption{\label{qiqc}
Channeltron amplification factor $A = \QC/\QI$
as a function of impact charge $\QI$ for all dust impacts 
detected between 2005 and 2007 with channeltron voltage set to HV = 4.  
The solid lines denote the sensitivity
threshold (lower left) and the saturation limit (upper right) of the channeltron. Squares
indicate dust particle impacts. The area of each square is proportional to 
the number of events included (the scaling of the squares is not the same 
as in earlier papers). The dotted horizontal line shows the mean value 
of the channeltron amplification A\,=\,1.57 
for ion impact charges $10^{-12}~\mathrm{C} <  \QI < 10^{-11}~\mathrm{C}$ 
(82 particles).
}
\end{figure}

The ratio of the channeltron charge $\QC$ and the ion collector
charge $\QI$ is a measure of the channeltron amplification A, which
in turn is an important parameter for dust 
impact identification (Paper~I).
In Figure~\ref{qiqc} we show the charge ratio $ \QC/\QI$ as a function of 
$ \QI$ for the 2005 to 2007 dust impacts with 
the channeltron high voltage set to 1250\,V (HV\,=\,4).
This diagram is directly comparable with similar 
diagrams in the previous Papers~III, V and VII for HV\,=\,3 and Paper~IX for
HV\,=\,4, respectively.

The mean amplification determined from particles with $10^{-12}{\rm\,C} \le \QI
\le 10^{-11}{\rm\,C}$ and HV~=~4 in the 2005 to 2007 interval is $\rm A \simeq 1.57$. This value is 
somewhat lower than the values derived from the first five years of the mission
(1990 to 1995; Papers~III and V) and comparable to the more recent
determinations (1996 to 2004; 
Papers~VII and IX). It indicates that a sufficiently high
channeltron amplification and stable instrument operation could be 
maintained with this higher voltage (HV~=~4) since 2000. 
It shows in particular that the channeltron degradation could
be mostly counterbalanced by increasing the high voltage by one digital
step. Much more severe electronics degradation was found for 
the Galileo dust detector during Galileo's orbital tour in the jovian 
system. It was likely related to the harsh radiation 
environment in the magnetosphere of the giant planet \citep{krueger2005a}. 

\begin{figure}
\vspace{-6cm}
\parbox{0.99\hsize}{
\hspace{-1cm}
\includegraphics[scale=0.8]{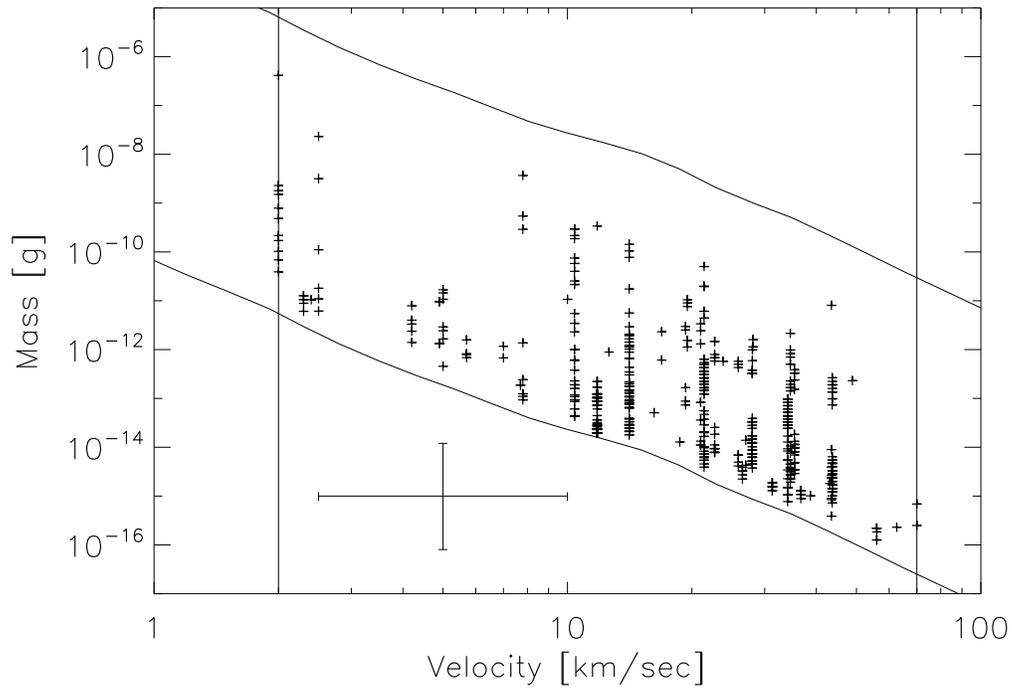}
}
\vspace{-9cm}
        \caption{\label{mass_speed}
Masses and impact velocities of all impacts recorded with the Ulysses sensor 
from 2005 to 2007. The lower and upper solid lines indicate the threshold and
the saturation limit of the detector, respectively, and the vertical lines 
indicate the calibrated velocity range. A sample error bar is shown that 
indicates a factor of 2 uncertainty for the velocity and a factor of 10 for 
the mass determination.
}
\end{figure}

In Figure~\ref{mass_speed} we show the masses and velocities of
all dust particles detected between 2005 and 2007. As in the earlier
periods before 2005, velocities occur over
the entire calibrated range from 2 to $\rm 70\,km\,s^{-1}$. The masses
vary over almost nine orders of magnitude from $\rm \sim 10^{-7}\,g$ to
$\rm 10^{-16}\,g$. The mean errors are a factor of 2 for the
velocity and a factor of 10 for the mass. The clustering
of the velocity values is due to discrete steps in the rise
time measurement but this quantisation is much smaller than the
velocity uncertainty. 
For many particles in the lowest two amplitude ranges (AR1 and
AR2) the velocity had to be computed from the ion charge signal
alone which leads to the vertical striping in the lower mass range in 
Figure~\ref{mass_speed} (most prominent above $\rm 10\,km\,s^{-1}$). In the
higher amplitude ranges the velocity could normally be calculated
from both the target and the ion charge signal, resulting in a 
more continuous distribution in the mass-velocity plane. Impact 
velocities below about $\rm 3\,km\,s^{-1}$ should be treated with caution
because anomalous impacts onto the sensor grids or structures 
other than the target generally lead to prolonged rise times and 
hence to unnaturally low impact velocities.

\section{Discussion} \label{discussion}

\begin{figure}
\vspace{-2cm}

\hspace{-1cm}
\parbox{0.99\hsize}{
\parbox{0.85\hsize}{
\includegraphics[scale=0.6]{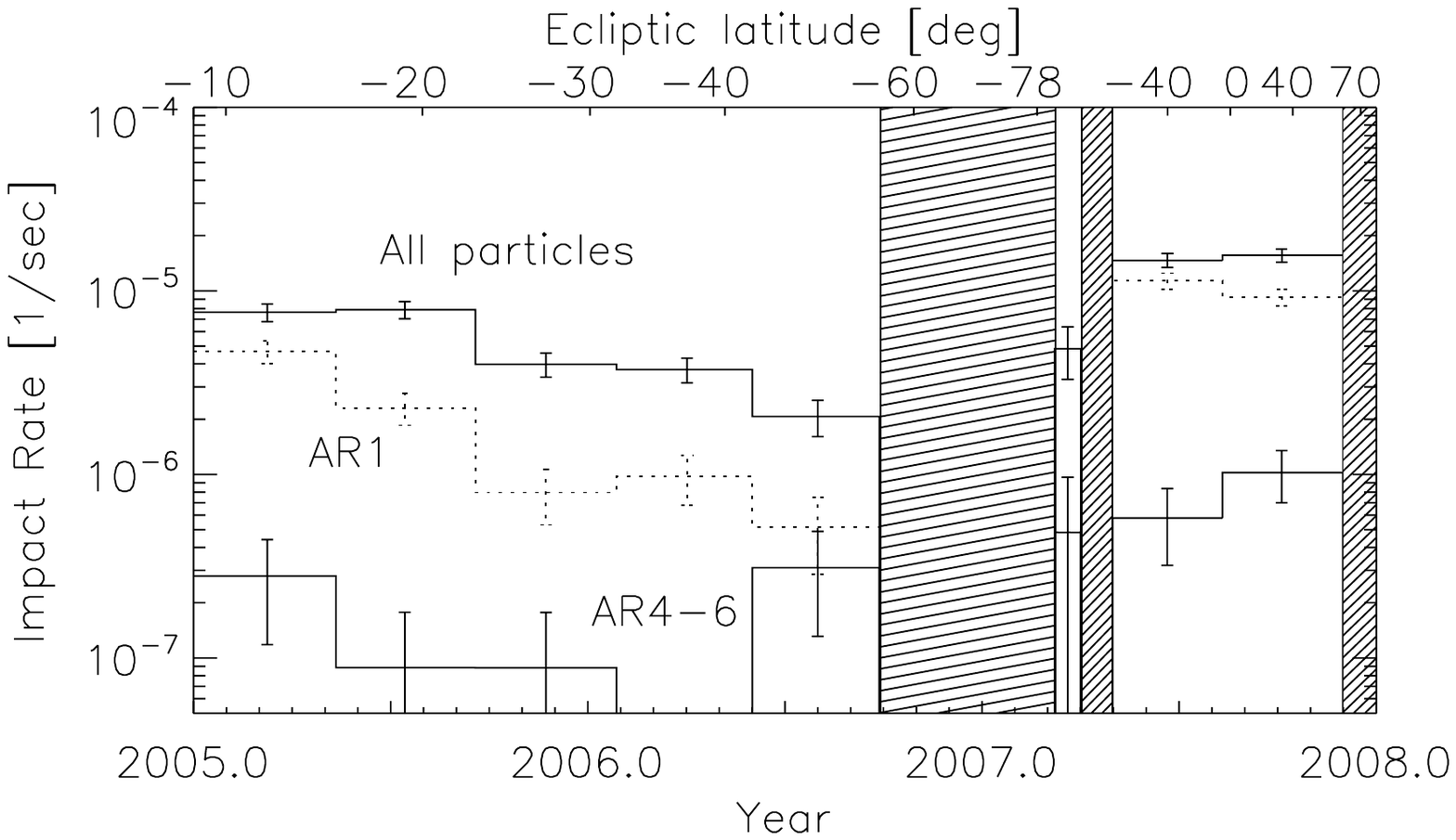}
}
\vspace{-10cm}

\parbox{0.85\hsize}{
\includegraphics[scale=0.6]{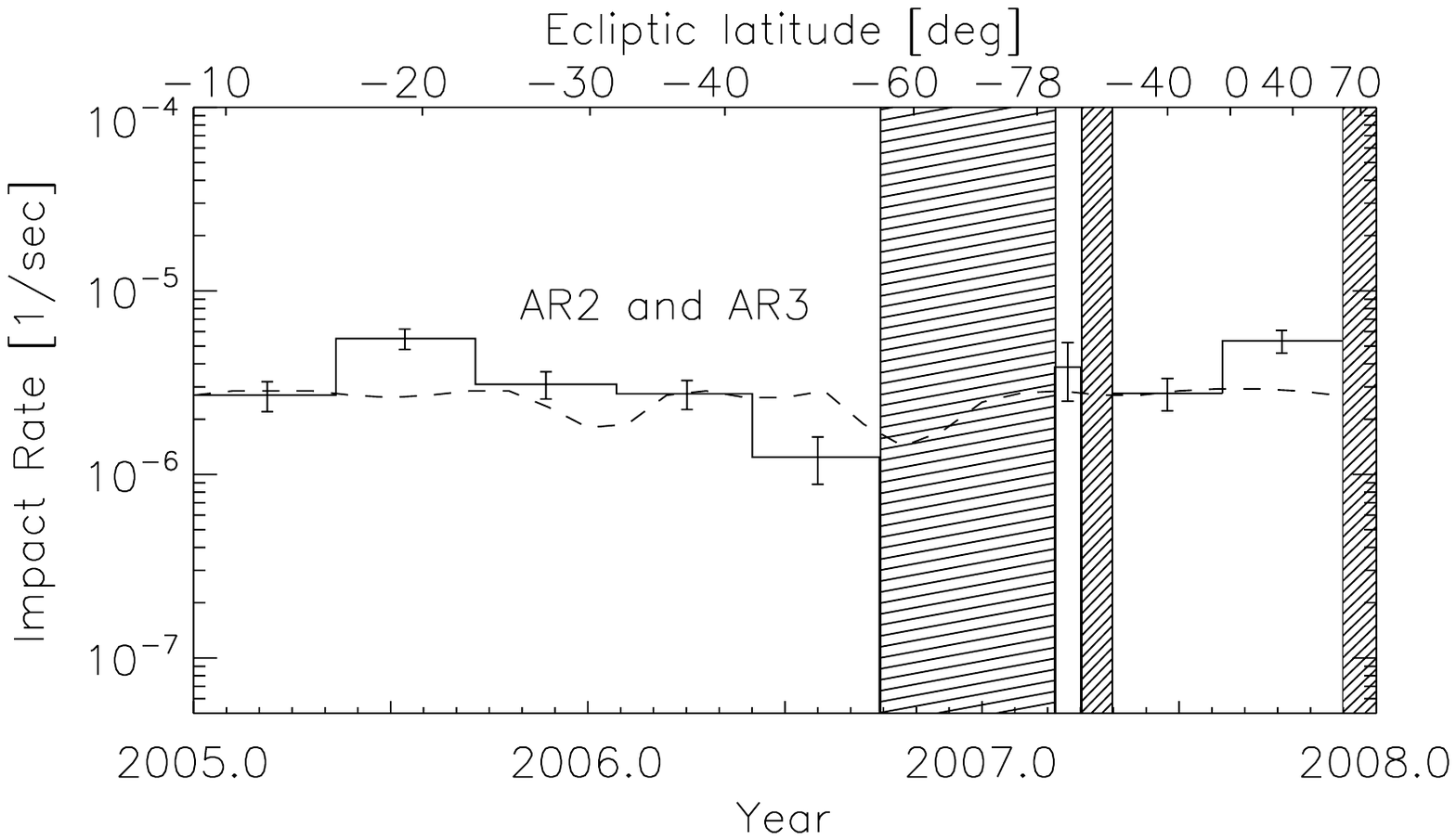}
}
\vspace{-7cm}
}

        \caption{\label{rate}
Impact rate of dust particles detected with the Ulysses dust sensor 
as a function of time. The ecliptic latitude of the spacecraft is
indicated at the top. Shaded areas indicate periods when the dust instrument 
was switched off. Upper panel: total impact rate (upper solid histogram),
impact rate of small particles (AR1, dotted histogram), and impact
rate of big particles (AR4 -- AR6, lower solid histogram).
Note that a rate of about $\rm 10^{-7}$ impacts per second 
is caused by a single dust impact in the averaging interval of about 
130 days.  An averaging interval of only 24 days had to be used for March 2007
because the instrument was switched off before and after this period. 
Lower panel: impact rate of 
intermediate size particles (AR2 and AR3, solid histogram). A model for the 
rate of interstellar 
particles assuming a constant flux was fit to the data and is superimposed as 
a dashed line. 
Vertical bars indicate the $\sqrt{n}$ statistical fluctuation. 
}
\end{figure}

In Figure~\ref{rate} we show the dust impact rate detected in various 
amplitude ranges together with the total impact rate summed over 
all amplitude ranges.  The highest overall
impact rate was recorded in 2007 when Ulysses was in the inner
solar system and relatively close to the ecliptic plane. 
It coincides with a peak in the rate of bigger particles in the three 
highest ion amplitude ranges (AR4 -- AR6). 
These impacts are attributed to interplanetary particles 
on low inclination orbits \citep{gruen1997a}. The majority of them are the 
impacts with $\QI > 10^{-10}~\mathrm{C}$ shown in Figure~\ref{nqi}.
The impact rate of intermediate sized particles in AR2 and AR3 showed 
relatively little variation. This size range is dominated by 
interstellar impactors. The details of the various dust 
populations are discussed further below.

\begin{figure}
\vspace{-1cm}
\parbox{0.99\hsize}{
\hspace{-1cm}
\includegraphics[scale=0.6]{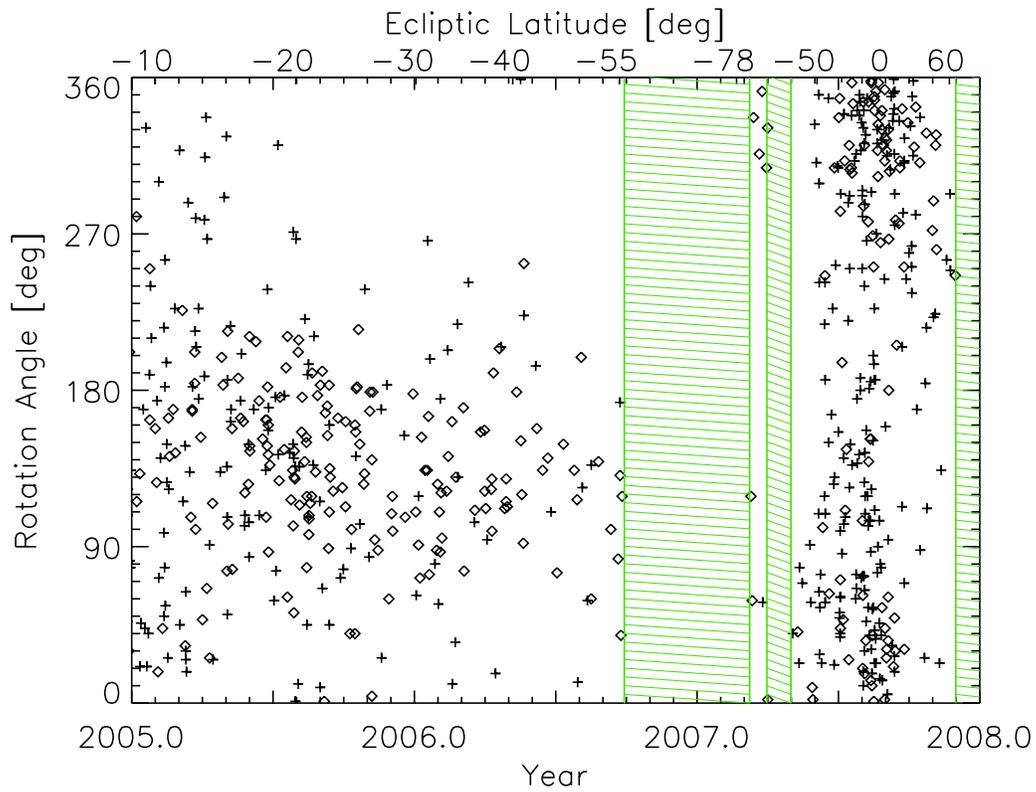}
}
        \caption{\label{rot_angle}
Rotation angle vs. time for all particles detected in the 2005-07
interval. Plus signs indicate particles 
with impact charge $\QI  < 10^{-13}$~C (AR1), diamonds those with
$ \QI  > 10^{-13}$~C (AR2 to AR6).  
Ulysses' ecliptic latitude is indicated at the top. 
}
\end{figure}

Figure~\ref{rot_angle} shows the sensor orientation at the time of a 
particle impact (rotation angle). 
The big particles (diamonds, 
impact charge $ \QI \rm \geq 8 \cdot 10^{-14}~C$ which roughly corresponds 
to AR2-6) are concentrated towards the upstream direction of interstellar 
helium \citep[cf. Figure~\ref{valeri}, bottom panel;][]{witte1996,witte2004a,witte2004b}. 
They have been detected with a relatively constant 
rate during the entire three-year period (Figure~\ref{rate}). 
The particles with the highest ion amplitude 
ranges (AR4 to AR6) are not distinguished in this diagram because 
they cannot be separated from interstellar particles by directional 
arguments alone. They have to be distinguished by other means
(e.~g. mass and speed). In addition, their total number is so small 
that they constitute only a small ''contamination`` of the interstellar 
particles in Figure~\ref{rot_angle}. 
In the ecliptic plane at 1.3~AU, 
however, interplanetary particle flux dominates over interstellar flux 
by a factor of about 3 (in number).

\subsection{Interstellar dust}
\label{interstellar_dust}

Interstellar particles move on hyperbolic trajectories through the 
solar system and approach Ulysses from the same direction as the 
interstellar gas \citep{gruen1994a,baguhl1995a,witte1996,witte2004a,witte2004b}. They 
can therefore be identified by their impact direction and
their impact speed. Earlier investigations showed that in the Ulysses and Galileo 
dust data sets the interstellar
impactors are mostly found in amplitude ranges AR2 and AR3. 

In the earlier mission from 1993 to 2004 the impact rate of interstellar grains 
(as derived from the AR2 and AR3 accumulators) varied by about 
a factor of 2.5: between 1993 and 1995 the rate was 
$\rm \sim 2 \cdot 10^{-6}\,s^{-1}$ (Paper~V) while in the 1996 to 1999 
interval it dropped to $\rm \sim 8 \cdot 10^{-7}\,s^{-1}$ (Paper~VII) and from  
2000 to 2004 the average impact rate was again
$\rm \sim 2 \cdot 10^{-6}\,s^{-1}$ (Paper~IX). In 2005/2006 the average 
impact rate of interstellar impactors was somewhat higher: 
$\rm \sim 3 \cdot 10^{-6}\,s^{-1}$ (bottom panel of Figure~\ref{rate}). 
In 2007, when Ulysses was in the inner solar system again, the impact rate derived 
from AR2 and AR3 was still higher. Here, however, the impact rate is not
due to interstellar particles alone because of a strong  
contribution from interplanetary impactors.

The dashed curve in Figure~\ref{rate} (bottom panel) shows the expected 
impact rate of interstellar particles assuming that they approach from the 
direction of interstellar helium \citep{witte1996,witte2004a,witte2004b} and 
that they move through the solar system on straight trajectories with 
a relative velocity of $26\ {\rm km}\ {\rm s}^{-1}$.
This assumption means dynamically that radiation pressure
cancels gravity for these particles ($\beta=1$) and that their 
Larmor radii are large compared with the dimension of the solar system. 
Both assumptions are reasonable for particles with masses between 
$10^{-13}$ and $10^{-12}\ {\rm g}$ which is the dominant 
size range measured for interstellar grains \citep{gruen1997a}.
The variation predicted 
by the model is caused by changes in the instrument's viewing 
direction with respect to the approach direction of the particles 
and changes in the relative velocity between the spacecraft and 
the particles. 
The dust particle flux is independent of heliocentric distance 
in this simple model, which gives relatively good agreement with the
observed impact rate. 

Ulysses has monitored the interstellar dust flow through the 
solar system for more than 15 years. This time period covers 
more than two and a half revolutions of the spacecraft about the
Sun through more than 2/3 of a complete 22-year solar cycle. Thus,
Ulysses measured interstellar dust during solar minimum and 
solar maximum conditions of the interplanetary magnetic field (IMF).
The interstellar dust flux modulation due to grain interaction with 
the magnetic field during
solar minimum could be well explained 
\citep{landgraf1998a,landgraf2000b,landgraf2003}. By taking into 
account the sensor side wall in the instrument field of view we could 
recently improve the flux determination \citep{altobelli2004a}. 

\begin{figure}

\hspace{-1cm}
\parbox{0.99\hsize}{
\parbox{0.65\hsize}{
\includegraphics[scale=0.35]{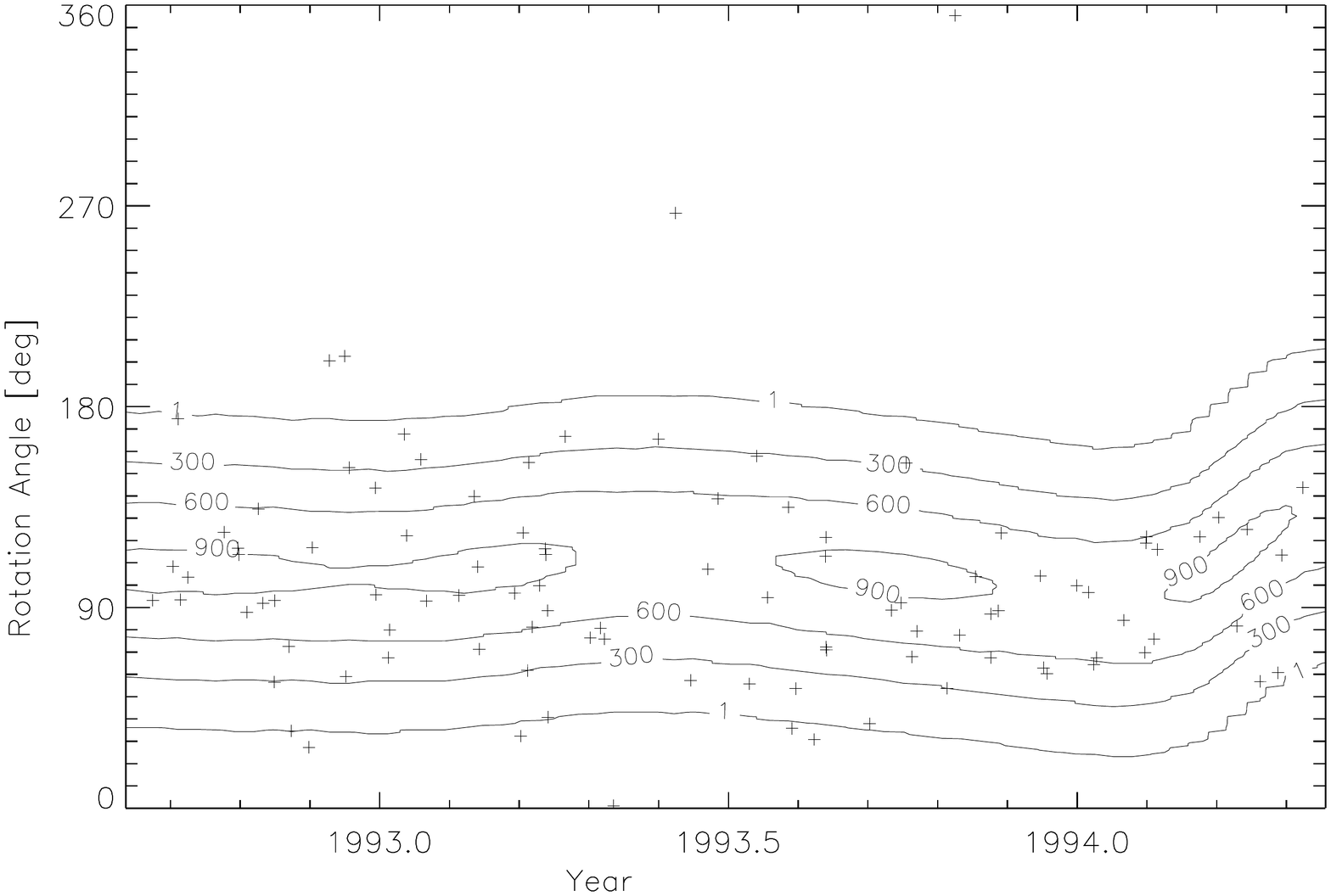}
}
\vspace{-1.5cm}

\parbox{0.65\hsize}{
\includegraphics[scale=0.35]{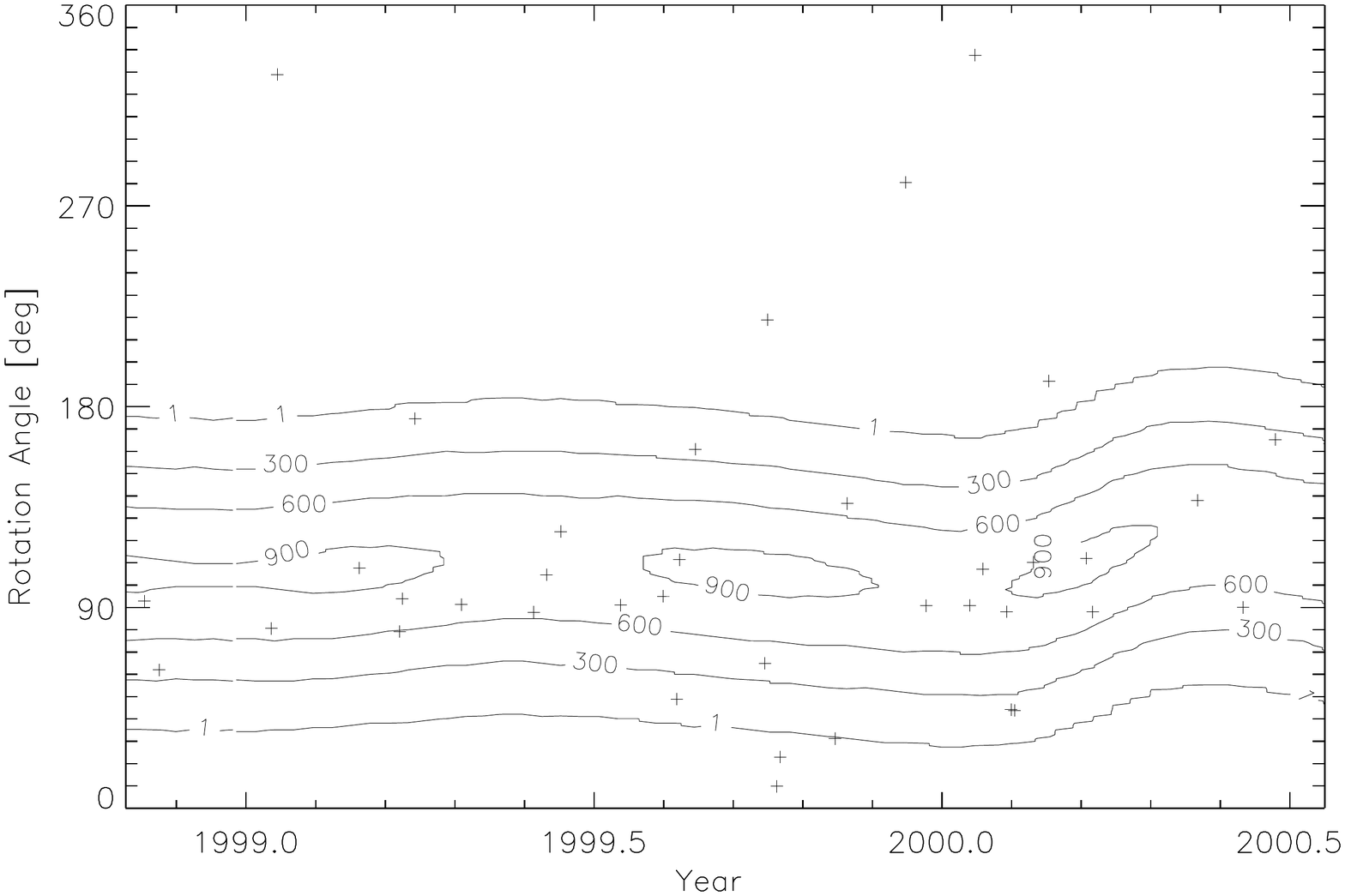}
}
\parbox{0.65\hsize}{
\vspace{-1.5cm}
\includegraphics[scale=0.35]{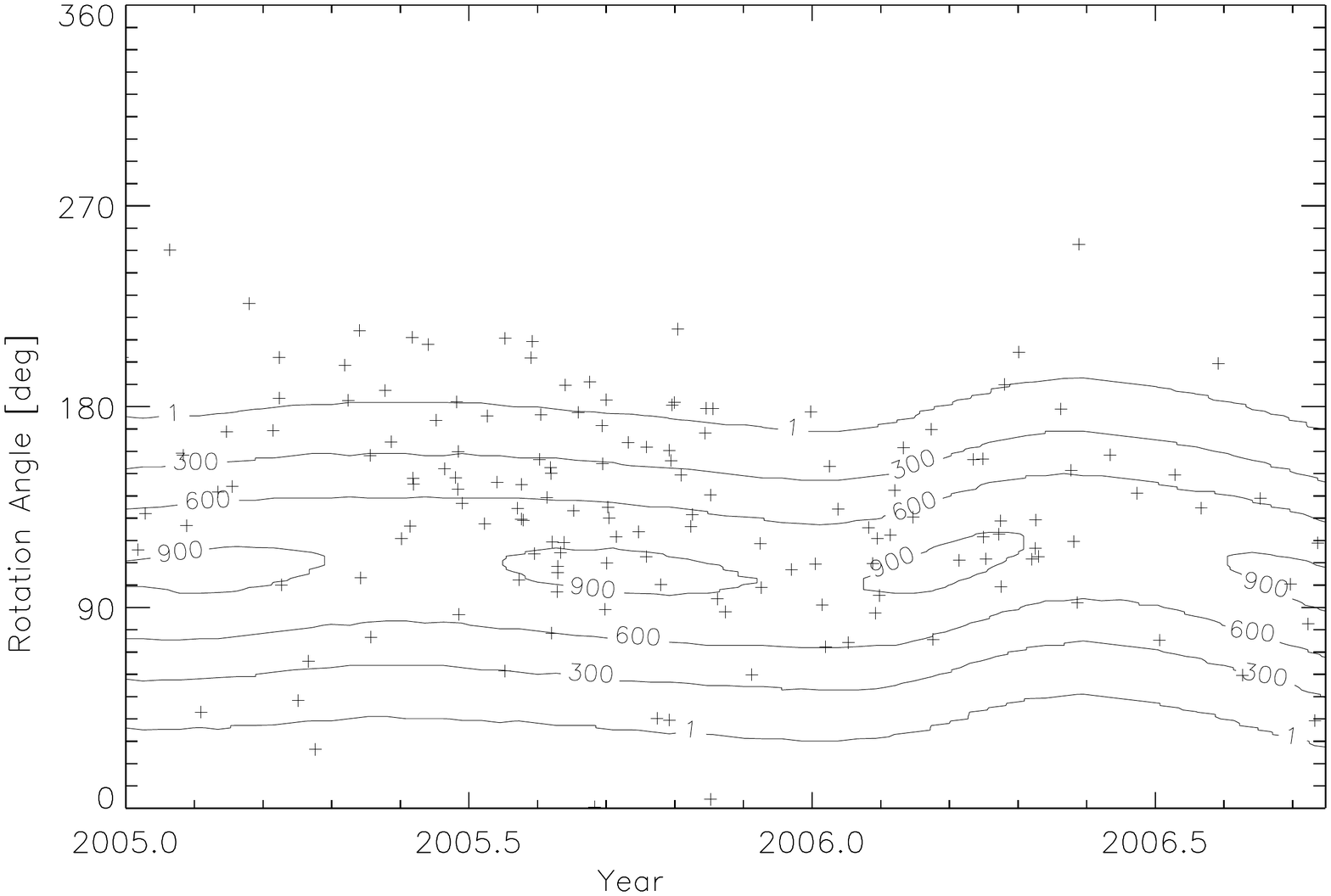}
}
}
\vspace{-1cm}
 \caption{\label{isd} 
Rotation angle vs. time for dust impacts detected during three time intervals when Ulysses 
was traversing the same portion of its trajectory between $-8.5^{\circ}$ and $-56^{\circ}$
ecliptic latitude. Contour lines show the effective sensor area for dust particles approaching from the
upstream direction of interstellar helium \citep{witte2004a,witte2004b}.
Here we show particles with impact charges $\mathrm{1.5 \cdot 10^{-13}~C < \QI < 10^{-11}~C}$ 
which approximately coincides with AR2-3.
}
\end{figure}

Ulysses provides the unique opportunity to compare repeated dust measurements
at the same locations in the solar system for different phases of the
solar cycle and the interplanetary magnetic field. 
In Figure~\ref{isd} we show the approach direction of interstellar grains for
three selected periods when Ulysses was between approximately -8 and $-56^{\circ}$
ecliptic latitude during approach to the inner solar system. During these
three intervals the dust detection geometry as indicated by the contour
lines was very similar so that we can compare them directly. At least
three differences are obvious between the three panels: 
\begin{itemize}
\item The approach direction of the majority of grains was compatible with the
flow direction of the interstellar helium gas as measured with Ulysses \citep{witte2004a,witte2004b}
during all three time intervals.
\item The dust flux varied by a factor of about two during the three  intervals,
the lowest flux being measured in the period 1999 to 2000, while the highest flux
occurred in 2005/2006 \citep{krueger2007b}.
\item A subset of the grains detected in 2005 shows a clear shift in the detected 
impact direction away from the ecliptic plane towards southern
ecliptic latitudes.
\end{itemize}

Our preliminary analysis 
indicates that this shift is about $\mathrm{30^{\circ}}$ away from 
the ecliptic plane towards southern ecliptic latitudes \citep{krueger2007b}. 
The reason for this shift remains mysterious. Whether it is connected
to a secondary stream of interstellar neutral atoms shifted from
the main neutral gas flow \citep{collier2004,wurz2004,nakagawa2006} is presently 
unclear.
Given, however, that the neutral gas stream is shifted along the ecliptic
plane while the shift in the dust flow is offset from the ecliptic,
a connection between both phenomena seems unlikely.

Even though Ulysses' position in the heliosphere and the dust detection conditions were 
very similar during all three time intervals considered in Figure~\ref{isd}, the configurations 
of the solar wind driven interplanetary magnetic field (IMF), which strongly affects the 
dynamics of the smallest grains, were completely different. We have to consider that the 
interstellar grains need approximately twenty years to travel from the heliospheric boundary 
to the inner solar system where they are detected by Ulysses. Thus, the effect of the IMF 
on the grain dynamics is the accumulated effect caused by the interaction with the IMF 
over several years: In the earlier time intervals (1993/1994 and 1999/2000) the grains had a 
recent dynamic history dominated by solar minimum conditions \citep{landgraf2000b}, while the 
grains detected during the third interval (2005/2006) had a recent history dominated by the much 
more disturbed solar maximum conditions of the IMF. During the solar maximum conditions the 
overall magnetic dipole field changed polarity. \citet{morfill1979b}  predicted that due to this 
effect in a 22-year cycle, small interstellar grains experience either focussing or defocusing 
conditions. During these times they are systematically deflected by the solar wind magnetic 
field either towards or away from the solar magnetic equator plane (close to the ecliptic plane). 
This latter configuration likely has a strong influence on the dust dynamics and the total 
interstellar flux in the inner heliosphere but it has not been modelled in detail.
An explanation of the grain interaction with the IMF at the recent solar 
maximum conditions is still pending.

Detailed modelling of the dynamics of the electrically 
charged dust grains in the heliosphere can give us information 
about the local interstellar environment of the solar system where the particles 
originate from. The models developed by \citet{landgraf2003} fit the observed flux variation by assuming 
a constant dust concentration in the local interstellar environment of our 
solar system. It implies that the local interstellar dust phase must be homogeneously distributed 
over length scales of $\mathrm{50\:AU}$, which is the distance  
travelled by the Sun during the measurement period of 
Ulysses from  1992 to  2002. This conclusion
is supported by the more recent Ulysses data until the end of 2004 
\citep{krueger2006b}. Our latest Ulysses dust data of 2005/2006, on the other hand, put a question 
mark onto this 
conclusion because if the observed shift in impact direction turns
out to be intrinsic, it would imply that this homogeneity breaks down on 
larger length scales.

\subsection{Interplanetary dust}

\label{ipd}

\begin{figure}
\hspace{-1cm}
\begin{turn}{270}
\includegraphics[scale=0.6]{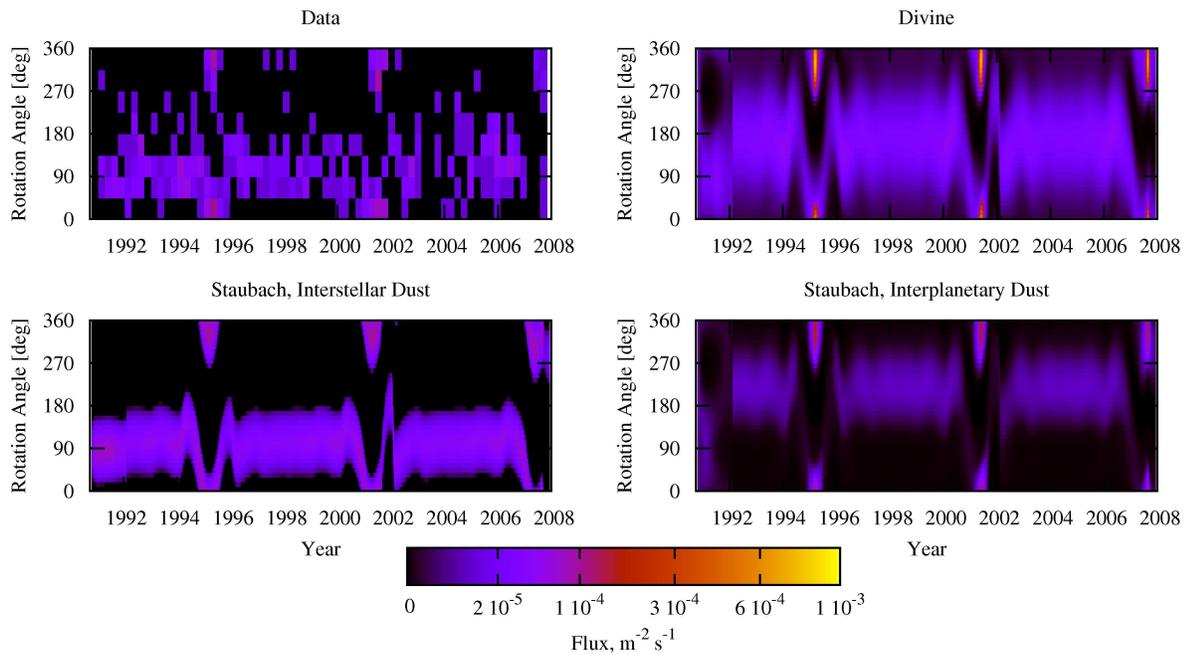}
\end{turn}
\vspace{-2cm}
        \caption{\label{valeri} 
Particle fluxes onto the Ulysses dust detector as seen in
the data (AR3 to AR6) and predicted by the meteoroid environment models by 
\citet{divine1993} and \citet{staubach1997}. The Jupiter encounter in 1992 is best
seen on the model plots as a sharp swing in directionality of impacts
caused by the rotation of spacecraft velocity relative to the giant
planet. The ecliptic plane crossings near the perihelia in 1995, 
2001 and 2007 are marked by the high impact rates due to the number density of
dust increase near the Sun and high speed of Ulysses relative to this
dust. As the spacecraft moved north at the times of the crossings, the
meteoroids came from the ecliptic north as well (rotation angle 0).
}
\end{figure}

Figure~\ref{valeri} shows the data (AR3 to AR6) from launch in 1990 until 
the end of the Ulysses mission in 2007 and compares them with two meteoroid environment 
models by 
\citet{divine1993} and \citet{staubach1997}. For most of the mission time,
excluding the ecliptic plane crossings in 1995, 2001 and 2007
the Divine model predicts impacts from a very broad range of directions,
with spin angles from $45^{\circ}$ to $300^{\circ}$. The impactors belong mostly
to the so-called "halo" population which was introduced in the model to
explain the Pioneer data \citep{divine1993}. The Ulysses directionality of the impacts
had not been available at the time of construction of the model, and
they are not well fit by the "halo" population. In contrast,
the \citet{staubach1997} model was fit to 5 years of Ulysses data from the craft's first
orbit about the Sun, taking the crucial directional information into account. It is
in better agreement with the data and has been confirmed by Ulysses' second and third orbits. 
It places most of the impacts into
the spin angle range from $30^{\circ}$ to $120^{\circ}$. These are due
to interstellar dust flowing through the solar system 
\citep{gruen1994a}.

One more observation from these plots is that both
the Divine and Staubach models predict
higher flux during the ecliptic plane crossings than the data permit.
The time dependence of the interstellar dust flux was asserted after the
meteoroid models under review had been constructed \citep{landgraf2000b}. 
However, the disagreement at the ecliptic plane crossings was not anticipated, since
almost all data incorporated in the models were taken from the ecliptic
plane. Possible explanations of the discrepancy are the roughness of
model fits as well as different representations of the data taken for
model adjustments and displayed in Figure~\ref{valeri}. 
While Figure~\ref{valeri} selects all 
impacts above a fixed threshold of charge released, the models were fitted 
using more uncertain inferred mass thresholds. The inference of mass is 
based on speed determination that is uncertain by a factor of 2.

\subsection{$\beta$-meteoroids}

When Ulysses was in the inner solar system in 1995 and in 2001,
maxima were evident in the impact rate of the smallest particles 
(AR1; Papers~V and IX). A similar maximum in the inner solar system 
occurred again in 2007 (Fig.~\ref{rate}; top panel).
Although in all three cases the maximum was reached 
during a short period around ecliptic plane crossing, many
particles were also detected at high ecliptic latitudes. These 
grains are attributed to a population of submicron-sized 
interplanetary particles whose dynamics is dominated by solar 
radiation pressure. They move on escape trajectories from the solar 
system \citep[$\beta$-meteoroids,][]{baguhl1995b,hamilton1996a}.
$\beta$-meteoroids were identified with Ulysses 
over the Sun's poles in 1994/95 \citep{wehry1999} and 2000/01 
\citep{wehry2004}. Due to the detection geometry, however, 
they were undetectable outside these periods, explaining the 
lower impact rates in AR1. $\beta$-meteoroids were
detectable again from mid-2006 until early-2008 \citep{wehry2004}. 
In independent potential detection of  $\beta$-meteoroids was
recently reported from the plasma wave instrument on board the 
STEREO spacecraft \citep{meyer-vernet2009}.

\citet{wehry1999} and \citet{wehry2004} identified a significant 
asymmetry in the flux of these particles between the northern and the
southern hemisphere from the first two heliocentric orbits of Ulysses,
the reason of which is presently unknown. 
A comprehensive analysis of these three data sets will be the 
subject of a future investigation and may reveal whether the 
asymmetry is connected with the solar cycle variation of the interplanetary 
magnetic field.

\section{Conclusions}                          \label{summary}

In this paper, the eleventh and final in a series of Ulysses and Galileo 
dust data papers, we present data from the Ulysses dust instrument for the 
period January 2005 to November 2007. In this time interval, starting
from a heliocentric distance of 5.3~AU close to aphelion, the
spacecraft approached the Sun, flew over the Sun's south pole, crossed the ecliptic 
plane at 1.4~AU heliocentric distance and reached a northern  
ecliptic latitude of $60^{\circ}$.  

A total number of 609 dust impacts were recorded during this period. 
Together with 6110 impacts recorded in interplanetary space and near Jupiter 
between Ulysses' launch in October 1990 and December 2004 
\citep{gruen1995c,krueger1999b,krueger2001b,krueger2006b}, the complete dust data set 
measured during the entire Ulysses mission 
consists of 6719 impacts. Given its temporal coverage and Ulysses'
unique orbital orientation, the Ulysses dust data set will be a treasure for
decades to come.

The total recorded dust impact rate dropped from an initial value of 
0.7 impacts per day in  2005 when Ulysses was at low ecliptic latitudes to a 
value of 0.3 impacts at higher latitudes. Most of these  
particles were of interstellar origin, in particular at higher ecliptic latitudes, 
a minor fraction being interplanetary particles. A maximum of 1.5 per day was 
measured in 2007 in the inner solar system; here the majority of the
grains were of interplanetary origin. 

The measurements from the entire Ulysses mission since launch
in 1990 are in disagreement with the interplanetary dust flux model by
\citet{divine1993}. Instead, they are well matched by  the model of
\citet{staubach1997} which was originally developed with a shorter 
data set from the first Ulysses orbit and with the 
dust measurements from Galileo's interplanetary cruise.

Noise tests performed regularly during the three years period revealed no 
degradation in the noise sensitivity of the dust instrument, and the
nominal instrument operational configuration remained unchanged during 
the entire period. In particular, no change in the channeltron high 
voltage setting was required. 

With Ulysses we had the unprecedented opportunity to measure dust in
interplanetary space for approximately 17 years. In particular, due to the unique 
orientation of Ulysses' orbital plane approximately perpendicular
to the flow direction of the interstellar dust through the solar system and 
the spacecraft's 6-year revolution 
period about the Sun, we obtained dust 
measurements from three traverses of the same spatial region at southern ecliptic
latitudes. These passes were separated by six years in time and were obtained at 
different phases of the
solar cycle. Variations in the interstellar dust flux by a factor of three are evident 
during the entire mission. Flux variations until 2002 have been explained by
the interaction of the dust grains with the time-varying interplanetary
magnetic field, while detailed modelling of the later data is still pending. 
The data obtained in 2005/2006 reveal an approximately $\mathrm{30^{\circ}}$ shift
in the approach direction of the grains away from the approach direction of the
interstellar helium gas. The reason for this shift remains mysterious and will 
be the subject of a future investigation. 

Even though this is the final paper in our serious of Ulysses dust data
papers published during the last 15 years, the evaluation of this unique data 
set is continuing. A list of specific open questions raised in this and earlier 
data papers includes: 
\begin{itemize}
\item {\em $\beta$-meteoroids:} The detailed evaluation of the measurements from Ulysses'
third solar orbit is still pending. Open questions include the measured 
north-south asymmetry in the measured flux and the heliocentric distance 
range where these grains are generated. More measurements from other spacecraft (e.g. STEREO)
would be advantageous to answer these questions.
\item {\em Comparison with Pioneer~10 and 11 measurements:} The flux of particles 
with $ m \rm \geq 10^{-9}\,g$ as measured 
with Ulysses is about a factor of five lower than expected from the 
Pioneer~10/11 measurements. Potential reasons for this
discrepancy are errors on the mass calibrations of the Ulysses and/or Pioneer detectors in 
this mass range, 
many of the Pioneer detections may 
not be due to actual meteoroid impacts, or the dust grains by Pioneer belong to a 
population of dust which could not or only partially be detected with Ulysses 
\citep{krueger1999b}. Measurements with New Horizons in the outer solar system
may shed new light onto this question. 
\item {\em Interstellar dust:} Earlier comprehensive investigations of the 
interstellar impactors were mostly performed in the late 1990s and 
relied upon the significantly smaller data set available at the time. In the meantime, 
until the end of the Ulysses mission, the interstellar dust data set has grown by at least 
a factor of two so that a complete re-analysis of the entire data set is  
worthwhile and can give
new insights into, e.g., the grain dynamics inside the heliosphere and into the conditions in the
local interstellar environment where these grains originate from. In particular,
the reason for the observed $\mathrm{30^{\circ}}$ shift remains an open question. 
\item {\em Interplanetary dust:} The interplanetary dust model by \citet{divine1993}
was developed before the Ulysses data became available and the model by \citet{staubach1997}
used only data from within the ecliptic plane and from a fraction of Ulysses' first 
heliocentric orbit. A new model is
presently being developed by \citet{dikarev2005} which will incorporate infrared observations 
of the zodiacal cloud by the COBE DIRBE 
instrument, in-situ flux measurements by the 
dust detectors on board Galileo and Ulysses, and the crater size distributions 
on lunar rock samples retrieved by the Apollo missions.
\end{itemize} 



\clearpage

\vspace{1cm}

{\bf Acknowledgements.}
We dedicate this work to the memory of Dietmar Linkert who passed away in 
spring 2009. He was Principal Engineer for space instruments at MPI f\"ur 
Kernphysik  including the
dust instruments flown on the HEOS-2, Helios, Galileo, Ulysses and Cassini missions. 
His friends and colleagues around the world appreciated his experience and 
sought his professional advice.
We thank the Ulysses project at ESA and NASA/JPL for effective and successful 
mission operations. 
This work has been supported by the Deutsches Zentrum f\"ur 
Luft- und Raumfahrt e.V. (DLR) under grants 50 0N 9107 and 50 QJ 9503. 
Support by Max-Planck-Institut f\"ur Kernphysik and Max-Planck-Institut f\"ur 
Sonnensystemforschung is also gratefully acknowledged.

\clearpage

\clearpage

\begin{sidewaystable}[h]
\caption{\label{papers}
Summary of Ulysses data papers, significant mission events and dust detector 
switch-off times. Jupiter distance $\mathrm{R_J=71492\,km}$.
}
\small
\begin{tabular}{llll}
&&&\\
\hline
\hline
Time Interval & Significant Mission Events                          & Dust Detector off  & Paper Number \\
\hline
1990 -- 1992  & Ulysses launch (6 Oct 1990),                              & Before 27 Oct 1990,          & III \citep{gruen1995c}\\
              & Jupiter flyby (8 Feb 1992, distance $\mathrm{6.3\,R_J}$)  & 14 Jun 1991 - 18 Jun 1991 & \\[1.5ex]
1993 -- 1995  & Maximum southern latitude $-79^{\circ}$ (3 Oct 1994),     & 8 Aug 1993 - 11 Aug 1993,  & V \citep{krueger1999b}    \\
              & Perihelion (12 Mar 1995),                                 & 27 Nov 1993 - 28 Nov 1993,  &    \\
              & Maximum northern latitude $79^{\circ}$ (19 Aug 1995)      & 10 Oct 1994 - 11 Oct 1994,  & \\ 
              &                                                           & 10 Dec 1995                & \\[1.5ex]
1996 -- 1999  & Aphelion (20 Apr 1998),                                   & 17 Aug 1996 - 18 Aug 1996,  & VII \citep{krueger2001b}\\
              &                                                           & 1 Apr 1997 - 2 Apr 1997,     & \\
              &                                                           & 15 Feb 1999 - 16 Feb 1999   & \\[1.5ex]
2000 -- 2004  & Maximum southern latitude $-80^{\circ}$ (27 Nov 2000),    & 26 Mar 2002 - 8 Apr 2002,   & IX \citep{krueger2006b}\\
              & Perihelion (23 May 2001),                                 & 1 Dec 2002 - 3 Jun 2003,    &       \\
              &  Maximum northern latitude $80^{\circ}$ (13 Oct 2001),     & 28 Jun 2003 - 22 Aug 2003,  & \\
              & Jupiter flyby (4 Feb 2004, distance 0.8~AU),              &  30 Nov 2003 - 2 Dec 2003  & \\
              &  Aphelion (30 Jun 2004)                                   &                             & \\[1.5ex]
2005 -- 2007  & Maximum southern latitude $-80^{\circ}$ (7 Feb 2007),        & 28 Sep 2006 - 9 Mar 2007,   & XI (this paper) \\
              & Perihelion (18 Aug 2007)                                  & 2 Apr 2007 - 30 Apr 2007,   & \\
              &                                                           & After 30 Nov 2007           & \\

\hline
\end{tabular}
\end{sidewaystable}

\clearpage

\thispagestyle{empty}

\begin{table}[htb]
\caption{\label{event_table} Ulysses mission and dust detector (GRU) 
configuration, tests and other events. 
Only selected events are given before 2005. See Section~\ref{mission} for 
details.
}
{\scriptsize
  \begin{tabular*}{15cm}{lccl}
   \hline
   \hline \\[-2.0ex]
Yr-day& 
Date&
Time& 
Event \\[0.7ex]
\hline \\[-2.0ex]
90-279 & 06.10.90&       & Ulysses launch \\
03-154& 03.06.03 & 03:38 &GRU 0.8 W heater on\\
04-182& 30.06.04&        &Ulysses aphelion passage (5.4~AU) \\
04-337& 02.12.04 & 11:00 &GRU nominal configuration: HV=4, EVD=C,I, SSEN=0001 \\
05-006& 06.01.05&  07:00 &GRU noise test\\
05-034& 03.02.05&  04:20 &GRU noise test\\
05-063& 04.03.05&  03:00 &GRU noise test\\
05-090& 31.03.05&  01:00 &GRU noise test\\
05-120& 30.04.05&  00:00 &GRU noise test\\
05-146& 26.05.05&  04:00 &GRU noise test\\
05-174& 23.06.05&  20:00 &GRU noise test\\
05-201& 20.07.05&  17:27 &GRU noise test\\
05-229& 17.08.05&  11:49 &GRU noise test\\
05-259& 16.09.05&  19:00 &GRU noise test\\
05-285& 12.10.05&  09:00 &GRU noise test\\
05-314& 10.11.05&  13:14 &GRU noise test\\
05-342& 08.12.05&  11:00 &GRU noise test\\
06-005& 05.01.06&  08:30 &GRU noise test\\
06-033& 02.02.06&  09:16 &GRU noise test\\
06-061& 02.03.06&  04:00 &GRU noise test\\
06-089& 30.03.06&  02:00 &GRU noise test\\
06-118& 28.04.06&  00:30 &GRU noise test\\
06-148& 28.05.06&  06:28 &GRU noise test\\
06-173& 22.06.06&  07:33 &GRU noise test\\
06-202& 21.07.06&  19:00 &GRU noise test\\
06-232& 20.08.06&  17:00 &GRU noise test\\
06-256& 13.09.06&  20:11 &GRU noise test\\
06-271& 28.09.06&  16:28 &GRU off (both heaters off) \\
07-038& 07.02.07&        &Ulysses maximum south solar latitude ($-79.7^{\circ}$)\\
07-067& 08.03.07&  18:08 &GRU 0.8 W heater on\\
07-067& 08.03.07&  18:28 &GRU 0.4 W heater on\\
07-068& 09.03.07&  16:44 &GRU 0.8 W heater off\\
07-068& 09.03.07&  17:31 &GRU on \\
07-069& 10.03.07&  07:54 &GRU nominal configuration \\
07-079& 20.03.07&  08:40 &GRU noise test\\
07-092& 02.04.07&  18:11 &GRU off (both heaters off) \\
07-120& 30.04.07&  01:23 &GRU 0.8 W heater on\\
07-120& 30.04.07&  01:33 &GRU 0.4 W heater on\\
07-120& 30.04.07&  21:37 &GRU 0.8 W heater off\\
07-120& 30.04.07&  21:38 &GRU on \\
07-121& 01.05.07&  04:07 &GRU nominal configuration \\
07-130& 10.05.07&  20:00 &GRU noise test\\
07-158& 07.06.07&  01:00 &GRU noise test\\
07-187& 06.07.07&  16:15 &GRU noise test\\
07-214& 02.08.07&  01:53 &GRU noise test\\
07-231& 18.08.07&        &Ulysses perihelion passage (1.39~AU) \\
07-232& 19.08.07&        &Ulysses ecliptic plane crossing \\
07-242& 30.08.07&  00:17 &GRU noise test\\
07-270& 27.09.07&  17:00 &GRU noise test\\
07-298& 25.10.07&  13:00 &GRU noise test\\
07-326& 22.11.07&  15:00 &GRU noise test\\
07-334& 30.11.07&  16:20 &GRU off (both heaters off) \\
08-014& 14.01.08&        &Ulysses maximum north solar latitude ($79.8^{\circ}$)\\
09-181& 30.06.09&        &Ulysses end of mission \\
\hline
\end{tabular*}\\[1.5ex]
Abbreviations used: 
HV: channeltron high voltage step; EVD: event definition,
ion- (I), channeltron- (C), or electron-channel (E); SSEN: detection thresholds,
ICP, CCP, ECP and PCP. 
Times when no data could be collected with the dust instrument:  
28 September 2006 to 9 March 2007, 2 April to 30 April 2007 and after 30 November 2007.
}
\end{table}

\thispagestyle{empty}

\clearpage

\thispagestyle{empty}
\begin{sidewaystable}
\tabcolsep1.5mm
\tiny
\vbox{
\hspace{-2cm}
\begin{minipage}[t]{22cm} 
{\normalsize Table~3. Overview of dust impacts detected with the Ulysses dust detector
between 1 January 2005 and 30 November 2007 as derived from the 
accumulators$^{\dagger}$.  Switch-on of the instrument  
is indicated by horizontal lines. 
The heliocentric distance R, 
the lengths of the time interval $\Delta $t (days) from the previous 
table entry, and the corresponding numbers of impacts are given for the 
24 accumulators. The accumulators are arranged with increasing signal 
amplitude ranges (AR), with four event classes for each amplitude 
range (CLN = 0,1,2,3); e.g.~AC31 means counter for AR = 1 and CLN = 3.  
The $\Delta $t in the first line (05-006)
is the time interval counted from the last entry in Table~2 in 
Paper~IX. The totals of  counted impacts$^{\dagger}$, of impacts with 
complete data, and of all events (noise plus impact events) 
for the entire period are given as well. }
\end{minipage}
}
\bigskip
\hspace{-2cm}
\begin{tabular}{|r|r|r|r|cccc|cccc|cccc|cccc|cccc|cccc|}
\hline
&&&&& &&&&& &&&&& &&&&& &&&&& &&\\
\mc{1}{|c|}{Date}&
\mc{1}{|c|}{Time}&
\mc{1}{|c|}{R}&
\mc{1}{|c|}{$\Delta $t}&
{\scriptsize AC}&{\scriptsize AC}&{\scriptsize AC}&{\scriptsize AC}&
{\scriptsize AC}&{\scriptsize AC}&{\scriptsize AC}&{\scriptsize AC}&
{\scriptsize AC}&{\scriptsize AC}&{\scriptsize AC}&{\scriptsize AC}&
{\scriptsize AC}&{\scriptsize AC}&{\scriptsize AC}&{\scriptsize AC}&
{\scriptsize AC}&{\scriptsize AC}&{\scriptsize AC}&{\scriptsize AC}&
{\scriptsize AC}&{\scriptsize AC}&{\scriptsize AC}&{\scriptsize AC}\\
&&[AU]&\mc{1}{c|}{[d]}&
\,\,01$^\dagger$&\,\,11$^\dagger$&21&31&
\,\,02$^\dagger$&12&22&32&
03&13&23&33&
04&14&24&34&
05&15&25&35&
06&16&26&36\\
&&&&& &&&&& &&&&& &&&&& &&&&& &&\\
\hline
&&&&& &&&&& &&&&& &&&&& &&&&& &&\\
05-006&11:45&      5.300& 7.561&-&-&-&-&-&-&-&1&-&-&-&-&-&-&-&1&$\ast$&-&-&-&$\ast$&-&-&-\\
05-039&23:36&      5.258& 33.49&-&1&11&2&-&-&5&2&-&-&-&-&-&-&-&-&$\ast$&-&-&-&$\ast$&-&-&-\\
05-061&19:36&      5.227& 21.83&-&3&8&3&-&-&-&2&-&-&-&1&-&-&1&-&$\ast$&-&-&-&$\ast$&-&-&-\\
05-089&09:46&      5.184& 27.59&-&1&11&3&-&1&5&2&-&-&-&1&-&-&-&-&$\ast$&-&-&-&$\ast$&-&-&-\\
05-104&17:01&      5.158& 15.30&-&-&4&2&-&-&3&-&-&-&-&-&-&-&-&1&$\ast$&-&-&-&$\ast$&-&-&-\\
 &&&&& &&&&& &&&&& &&&&& &&&&& &&\\
05-140&22:27&      5.089& 36.22&-&5&6&1&-&-&5&2&-&-&-&-&-&-&-&1&$\ast$&-&-&-&$\ast$&-&-&-\\
05-177&01:56&      5.013& 36.14&-&5&4&2&-&-&9&9&-&-&-&-&-&-&-&-&$\ast$&-&-&-&$\ast$&-&-&-\\
05-204&13:40&      4.949& 27.48&-&2&5&1&-&-&2&2&-&-&-&3&-&-&-&-&$\ast$&-&-&-&$\ast$&-&-&-\\
05-227&18:13&      4.891& 23.18&-&4&5&2&-&-&3&7&-&-&1&6&-&-&-&-&$\ast$&-&-&-&$\ast$&-&-&-\\
05-250&17:23&      4.831& 22.96&-&1&6&-&-&1&1&7&-&-&-&2&-&-&-&1&$\ast$&-&-&-&$\ast$&-&-&-\\
 &&&&& &&&&& &&&&& &&&&& &&&&& &&\\
05-284&14:06&      4.734& 33.86&-&2&3&-&-&-&2&4&-&-&1&7&-&-&-&-&$\ast$&-&-&-&$\ast$&-&-&-\\
05-310&21:37&      4.654& 26.31&-&1&3&-&-&-&5&5&-&-&-&3&-&-&-&-&$\ast$&-&-&-&$\ast$&-&-&-\\
05-337&10:38&      4.568& 26.54&-&3&-&-&-&-&2&2&-&-&-&1&-&-&-&1&$\ast$&-&-&-&$\ast$&-&-&-\\
05-364&06:07&      4.475& 26.81&-&1&-&-&-&-&1&1&-&-&-&1&-&-&-&-&$\ast$&-&-&-&$\ast$&-&-&-\\
06-028&18:17&      4.368& 29.50&-&2&2&1&-&1&3&4&-&-&-&1&-&-&-&-&$\ast$&-&-&-&$\ast$&-&-&-\\
 &&&&& &&&&& &&&&& &&&&& &&&&& &&\\
06-052&22:35&      4.274& 24.17&-&2&3&-&-&-&2&4&-&-&-&2&-&-&-&-&$\ast$&-&-&-&$\ast$&-&-&-\\
06-078&05:59&      4.172& 25.30&-&2&2&-&-&-&1&-&-&-&-&3&-&-&-&-&$\ast$&-&-&-&$\ast$&-&-&-\\
06-100&06:53&      4.078& 22.03&-&-&1&-&-&-&3&1&-&-&1&1&-&-&-&-&$\ast$&-&-&-&$\ast$&-&-&-\\
06-132&07:12&      3.935& 32.01&-&-&2&-&-&-&2&1&-&-&-&5&-&-&-&-&$\ast$&-&-&-&$\ast$&-&-&-\\
06-165&22:45&      3.776& 33.64&-&1&2&1&-&-&1&2&-&-&-&2&-&-&-&1&$\ast$&-&-&-&$\ast$&-&-&-\\
 &&&&& &&&&& &&&&& &&&&& &&&&& &&\\
06-192&21:07&      3.642& 26.93&-&-&1&-&-&-&1&1&-&-&-&1&-&-&-&-&$\ast$&-&-&-&$\ast$&-&-&-\\
06-217&08:19&      3.514& 24.46&-&1&1&-&-&-&-&1&-&-&-&1&-&-&-&1&$\ast$&-&-&-&$\ast$&-&-&-\\
06-238&08:40&      3.400& 21.01&-&-&2&-&-&-&-&1&-&-&-&1&-&-&-&-&$\ast$&-&-&-&$\ast$&-&-&-\\
06-271&16:30&      3.211& 33.32&-&-&1&-&-&-&1&2&-&-&-&1&-&-&-&1&$\ast$&-&-&-&$\ast$&-&-&-\\
07-069&07:54&      2.163& 162.6&\mc{4}{|c|}{---------------------}&\mc{4}{|c|}{---------------------}&\mc{
4}{|c|}{---------------------}&\mc{4}{|c|}{---------------------}&\mc{4}{|c|}{---------------------}&\mc{4
}{|c|}{---------------------}\\
 &&&&& &&&&& &&&&& &&&&& &&&&& &&\\
07-092&18:15&      2.004& 23.43&-&1&-&-&-&1&2&2&-&-&1&1&-&-&-&-&$\ast$&-&-&1&$\ast$&-&-&-\\
07-121&04:07&      1.818& 28.41&\mc{4}{|c|}{---------------------}&\mc{4}{|c|}{---------------------}&\mc{
4}{|c|}{---------------------}&\mc{4}{|c|}{---------------------}&\mc{4}{|c|}{---------------------}&\mc{4
}{|c|}{---------------------}\\
07-152&04:50&      1.634& 31.02&-&1&6&-&-&1&-&1&-&-&1&-&-&-&-&-&$\ast$&-&-&-&$\ast$&-&-&-\\
07-182&21:34&      1.489& 30.69&-&13&10&4&-&-&-&1&-&-&-&2&-&-&1&-&$\ast$&-&-&-&$\ast$&-&-&-\\
07-200&08:56&      1.433& 17.47&2&12&6&1&-&-&8&5&-&-&-&1&-&-&-&1&$\ast$&-&-&-&$\ast$&-&-&-\\
 &&&&& &&&&& &&&&& &&&&& &&&&& &&\\
07-222&17:12&      1.396& 22.34&2&25&17&3&-&1&3&2&-&-&3&-&-&-&1&3&$\ast$&-&-&-&$\ast$&-&-&-\\
07-243&23:21&      1.402& 21.25&3&25&8&1&-&3&9&2&-&-&3&5&-&-&1&2&$\ast$&-&2&1&$\ast$&-&1&-\\
07-266&13:56&      1.452& 22.60&-&5&13&4&-&2&6&-&-&-&4&5&-&-&-&2&$\ast$&-&-&-&$\ast$&-&-&-\\
07-288&22:33&      1.538& 22.35&-&8&9&-&-&1&1&-&-&-&1&2&-&-&-&1&$\ast$&-&-&-&$\ast$&-&-&-\\
07-309&22:37&      1.644& 21.00&-&1&5&-&-&-&2&-&-&-&-&2&-&-&-&2&$\ast$&-&-&-&$\ast$&-&-&-\\
 &&&&& &&&&& &&&&& &&&&& &&&&& &&\\
\hline
\end{tabular}
\end{sidewaystable}

\clearpage

\thispagestyle{empty}
\begin{sidewaystable}
\tabcolsep1.5mm
\tiny
\vbox{
\hspace{-2cm}
\begin{minipage}[t]{22cm}   
{\normalsize Table~3 continued.}
\end{minipage}
}
\bigskip
\hspace{-2cm}
\begin{tabular}{|r|r|r|r|cccc|cccc|cccc|cccc|cccc|cccc|}
\hline
&&&&& &&&&& &&&&& &&&&& &&&&& &&\\
\mc{1}{|c|}{Date}&
\mc{1}{|c|}{Time}&
\mc{1}{|c|}{R}&
\mc{1}{|c|}{$\Delta $t}&
{\scriptsize AC}&{\scriptsize AC}&{\scriptsize AC}&{\scriptsize AC}&
{\scriptsize AC}&{\scriptsize AC}&{\scriptsize AC}&{\scriptsize AC}&
{\scriptsize AC}&{\scriptsize AC}&{\scriptsize AC}&{\scriptsize AC}&
{\scriptsize AC}&{\scriptsize AC}&{\scriptsize AC}&{\scriptsize AC}&
{\scriptsize AC}&{\scriptsize AC}&{\scriptsize AC}&{\scriptsize AC}&
{\scriptsize AC}&{\scriptsize AC}&{\scriptsize AC}&{\scriptsize AC}\\
&&[AU]&\mc{1}{c|}{[d]}&
\,\,01$^\dagger$&\,\,11$^\dagger$&21&31&
\,\,02$^\dagger$&12&22&32&
03&13&23&33&
04&14&24&34&
05&15&25&35&
06&16&26&36\\
&&&&& &&&&& &&&&& &&&&& &&&&& &&\\
\hline
&&&&& &&&&& &&&&& &&&&& &&&&& &&\\
07-334&16:30&      1.790& 24.74&-&3&2&-&-&-&-&-&-&-&1&-&-&-&-&-&$\ast$&-&-&-&$\ast$&-&-&-\\
\hline 
\mc{4}{|l|}{}&& &&&&& &&&&& &&&&& &&&&& &&\\
\mc{4}{|l|}{Impacts (counted)}&7$^{\dagger}$ &131$^{\dagger}$ &159&31&0$^{\dagger}$ &12&88&76&0&0&17&61&0&0&4&19&$\ast$&
0&2&2&$\ast$&0&1&0\\[1.5ex]
\mc{4}{|l|}{Impacts (complete data)}&7&131&158&31&0&12&88&76&0&0&17&61&0&0&4&19&$\ast$&0&2&2&$\ast$&0&1&0\\[1.5ex]
\mc{4}{|l|}{All events(complete data)}&6243&132&159&31&120&12&88&76&2&0&17&61&1&0&4&19&$\ast$&0&2&2&$\ast$&0&1&0\\[1.5ex]
\hline
\end{tabular}
\\
\parbox{17cm}{
\vspace{0.4cm}
\hspace{-2cm}
$\dagger$: {\small Entries for AC01, AC11 and AC02 are the number of impacts 
with complete data. Due to the noise contamination of these three categories 
the number of impacts cannot be determined from the accumulators. The method 
to separate dust impacts from noise events in these three categories has been given 
by Baguhl et al. (1993). \\
$\ast$: No entries are given for AC05 and AC06 because they count the overflows
of accumulators AC21 and AC31 since the reprogramming in December 2004 (Paper~IX).}}
\end{sidewaystable}

\clearpage

{\renewcommand{\baselinestretch}{0.8} 

\thispagestyle{empty}
 \begin{sidewaystable}
\tabcolsep1.2mm
\tiny
\vbox{
\hspace{-2cm}
\begin{minipage}[t]{21.5cm}      
 Table~4.
 Raw data: No., impact time, CLN, AR, SEC, IA, EA, CA, IT, ET, 
 EIT, EIC, ICC, PA, PET, EVD, ICP, ECP, CCP, PCP, HV and 
 evaluated data: R, LON, LAT, $D_{\rm Jup}$, 
 rotation angle 
(ROT), instr. pointing (S$_{LON}$, S$_{LAT}$), impact speed ($v$, in $\rm km\,s^{-1}$), 
 speed error factor (VEF), mass ($m$, in grams) and mass error factor (MEF).
\end{minipage}
}
\bigskip

\hspace{-2cm}
\begin{tabular*}{23cm} {@{\extracolsep{\fill}}                                  
rrrrrrrrrrrrrrrcrrrrcrrrrrrrrrrrrrrrrrrrrr}                                     
 \mc{1}{c}{No.}&               \mc{1}{c}{IMP. DATE}&         \mc{1}{c}{C}&                 \mc{1}{c}{AR}&                
 \mc{1}{c}{S}&                 \mc{1}{c}{IA}&                \mc{1}{c}{EA}&                \mc{1}{c}{CA}&                
 \mc{1}{c}{IT}&                \mc{1}{c}{ET}&                \mc{1}{c}{E}&                 \mc{1}{c}{E}&                 
 \mc{1}{c}{I}&                 \mc{1}{c}{PA}&                \mc{1}{c}{P}&                 \mc{1}{c}{E}&                 
 \mc{1}{c}{I}&                 \mc{1}{c}{E}&                 \mc{1}{c}{C}&                 \mc{1}{c}{P}&                 
 \mc{1}{c}{HV}&                \mc{1}{c}{R}&                 \mc{1}{c}{LON}&               \mc{1}{c}{LAT}&               
 \mc{1}{c}{D$_{\rm Jup}$}&     \mc{1}{c}{ROT}&               \mc{1}{c}{S$_{\rm LON}$}&     \mc{1}{c}{S$_{\rm LAT}$}&     
 \mc{1}{c}{V}&                 \mc{1}{c}{VEF}&               \mc{1}{c}{M}&                 \mc{1}{c}{MEF}\\
\end{tabular*} \end{sidewaystable}                                                     
\clearpage
\thispagestyle{empty}
\begin{sidewaystable}                                                                   
\tiny \hspace{-2cm}
\tabcolsep1.2mm
\begin{tabular*}{23cm} {@{\extracolsep{\fill}}                                  
rrrrrrrrrrrrrrrcrrrrcrrrrrrrrrrrrrrrrrrrrr}                                     
 \mc{1}{c}{No.}&               \mc{1}{c}{IMP. DATE}&         \mc{1}{c}{C}&                 \mc{1}{c}{AR}&                
 \mc{1}{c}{S}&                 \mc{1}{c}{IA}&                \mc{1}{c}{EA}&                \mc{1}{c}{CA}&                
 \mc{1}{c}{IT}&                \mc{1}{c}{ET}&                \mc{1}{c}{E}&                 \mc{1}{c}{E}&                 
 \mc{1}{c}{I}&                 \mc{1}{c}{PA}&                \mc{1}{c}{P}&                 \mc{1}{c}{E}&                 
 \mc{1}{c}{I}&                 \mc{1}{c}{E}&                 \mc{1}{c}{C}&                 \mc{1}{c}{P}&                 
 \mc{1}{c}{HV}&                \mc{1}{c}{R}&                 \mc{1}{c}{LON}&               \mc{1}{c}{LAT}&               
 \mc{1}{c}{D$_{\rm Jup}$}&     \mc{1}{c}{ROT}&               \mc{1}{c}{S$_{\rm LON}$}&     \mc{1}{c}{S$_{\rm LAT}$}&     
 \mc{1}{c}{V}&                 \mc{1}{c}{VEF}&               \mc{1}{c}{M}&                 \mc{1}{c}{MEF}\\
\end{tabular*} \end{sidewaystable}                                                     
\clearpage                                                                      
\thispagestyle{empty}
\begin{sidewaystable}                                                                   \
\tiny \hspace{-2cm}
\tabcolsep1.2mm
\begin{tabular*}{23cm} {@{\extracolsep{\fill}}                                  
rrrrrrrrrrrrrrrcrrrrcrrrrrrrrrrrrrrrrrrrrr}                                     
 \mc{1}{c}{No.}&               \mc{1}{c}{IMP. DATE}&         \mc{1}{c}{C}&                 \mc{1}{c}{AR}&                
 \mc{1}{c}{S}&                 \mc{1}{c}{IA}&                \mc{1}{c}{EA}&                \mc{1}{c}{CA}&                
 \mc{1}{c}{IT}&                \mc{1}{c}{ET}&                \mc{1}{c}{E}&                 \mc{1}{c}{E}&                 
 \mc{1}{c}{I}&                 \mc{1}{c}{PA}&                \mc{1}{c}{P}&                 \mc{1}{c}{E}&                 
 \mc{1}{c}{I}&                 \mc{1}{c}{E}&                 \mc{1}{c}{C}&                 \mc{1}{c}{P}&                 
 \mc{1}{c}{HV}&                \mc{1}{c}{R}&                 \mc{1}{c}{LON}&               \mc{1}{c}{LAT}&               
 \mc{1}{c}{D$_{\rm Jup}$}&     \mc{1}{c}{ROT}&               \mc{1}{c}{S$_{\rm LON}$}&     \mc{1}{c}{S$_{\rm LAT}$}&     
 \mc{1}{c}{V}&                 \mc{1}{c}{VEF}&               \mc{1}{c}{M}&                 \mc{1}{c}{MEF}\\
\end{tabular*} \end{sidewaystable}                                                     
\clearpage                                                                      
\thispagestyle{empty}
\begin{sidewaystable}                                                                   \
\tiny \hspace{-2cm}
\tabcolsep1.2mm
\begin{tabular*}{23cm} {@{\extracolsep{\fill}}                                  
rrrrrrrrrrrrrrrcrrrrcrrrrrrrrrrrrrrrrrrrrr}                                     
 \mc{1}{c}{No.}&               \mc{1}{c}{IMP. DATE}&         \mc{1}{c}{C}&                 \mc{1}{c}{AR}&                
 \mc{1}{c}{S}&                 \mc{1}{c}{IA}&                \mc{1}{c}{EA}&                \mc{1}{c}{CA}&                
 \mc{1}{c}{IT}&                \mc{1}{c}{ET}&                \mc{1}{c}{E}&                 \mc{1}{c}{E}&                 
 \mc{1}{c}{I}&                 \mc{1}{c}{PA}&                \mc{1}{c}{P}&                 \mc{1}{c}{E}&                 
 \mc{1}{c}{I}&                 \mc{1}{c}{E}&                 \mc{1}{c}{C}&                 \mc{1}{c}{P}&                 
 \mc{1}{c}{HV}&                \mc{1}{c}{R}&                 \mc{1}{c}{LON}&               \mc{1}{c}{LAT}&               
 \mc{1}{c}{D$_{\rm Jup}$}&     \mc{1}{c}{ROT}&               \mc{1}{c}{S$_{\rm LON}$}&     \mc{1}{c}{S$_{\rm LAT}$}&     
 \mc{1}{c}{V}&                 \mc{1}{c}{VEF}&               \mc{1}{c}{M}&                 \mc{1}{c}{MEF}\\
\end{tabular*} \end{sidewaystable}                                                     
\clearpage                                                                      
\thispagestyle{empty}
\begin{sidewaystable}                                                                   \
\tiny \hspace{-2cm}
\tabcolsep1.2mm
\begin{tabular*}{23cm} {@{\extracolsep{\fill}}                                  
rrrrrrrrrrrrrrrcrrrrcrrrrrrrrrrrrrrrrrrrrr}                                     
 \mc{1}{c}{No.}&               \mc{1}{c}{IMP. DATE}&         \mc{1}{c}{C}&                 \mc{1}{c}{AR}&                
 \mc{1}{c}{S}&                 \mc{1}{c}{IA}&                \mc{1}{c}{EA}&                \mc{1}{c}{CA}&                
 \mc{1}{c}{IT}&                \mc{1}{c}{ET}&                \mc{1}{c}{E}&                 \mc{1}{c}{E}&                 
 \mc{1}{c}{I}&                 \mc{1}{c}{PA}&                \mc{1}{c}{P}&                 \mc{1}{c}{E}&                 
 \mc{1}{c}{I}&                 \mc{1}{c}{E}&                 \mc{1}{c}{C}&                 \mc{1}{c}{P}&                 
 \mc{1}{c}{HV}&                \mc{1}{c}{R}&                 \mc{1}{c}{LON}&               \mc{1}{c}{LAT}&               
 \mc{1}{c}{D$_{\rm Jup}$}&     \mc{1}{c}{ROT}&               \mc{1}{c}{S$_{\rm LON}$}&     \mc{1}{c}{S$_{\rm LAT}$}&     
 \mc{1}{c}{V}&                 \mc{1}{c}{VEF}&               \mc{1}{c}{M}&                 \mc{1}{c}{MEF}\\
\end{tabular*} \end{sidewaystable}                                                     
\clearpage                                                                      
\thispagestyle{empty}
\begin{sidewaystable}                                                                   \
\tiny \hspace{-2cm}
\tabcolsep1.2mm
\begin{tabular*}{23cm} {@{\extracolsep{\fill}}                                  
rrrrrrrrrrrrrrrcrrrrcrrrrrrrrrrrrrrrrrrrrr}                                     
 \mc{1}{c}{No.}&               \mc{1}{c}{IMP. DATE}&         \mc{1}{c}{C}&                 \mc{1}{c}{AR}&                
 \mc{1}{c}{S}&                 \mc{1}{c}{IA}&                \mc{1}{c}{EA}&                \mc{1}{c}{CA}&                
 \mc{1}{c}{IT}&                \mc{1}{c}{ET}&                \mc{1}{c}{E}&                 \mc{1}{c}{E}&                 
 \mc{1}{c}{I}&                 \mc{1}{c}{PA}&                \mc{1}{c}{P}&                 \mc{1}{c}{E}&                 
 \mc{1}{c}{I}&                 \mc{1}{c}{E}&                 \mc{1}{c}{C}&                 \mc{1}{c}{P}&                 
 \mc{1}{c}{HV}&                \mc{1}{c}{R}&                 \mc{1}{c}{LON}&               \mc{1}{c}{LAT}&               
 \mc{1}{c}{D$_{\rm Jup}$}&     \mc{1}{c}{ROT}&               \mc{1}{c}{S$_{\rm LON}$}&     \mc{1}{c}{S$_{\rm LAT}$}&     
 \mc{1}{c}{V}&                 \mc{1}{c}{VEF}&               \mc{1}{c}{M}&                 \mc{1}{c}{MEF}\\
\end{tabular*} \end{sidewaystable}                                                     
\clearpage                                                                      
\thispagestyle{empty}
\begin{sidewaystable}                                                                   \
\tiny \hspace{-2cm}
\tabcolsep1.2mm
\begin{tabular*}{23cm} {@{\extracolsep{\fill}}                                  
rrrrrrrrrrrrrrrcrrrrcrrrrrrrrrrrrrrrrrrrrr}                                     
 \mc{1}{c}{No.}&               \mc{1}{c}{IMP. DATE}&         \mc{1}{c}{C}&                 \mc{1}{c}{AR}&                
 \mc{1}{c}{S}&                 \mc{1}{c}{IA}&                \mc{1}{c}{EA}&                \mc{1}{c}{CA}&                
 \mc{1}{c}{IT}&                \mc{1}{c}{ET}&                \mc{1}{c}{E}&                 \mc{1}{c}{E}&                 
 \mc{1}{c}{I}&                 \mc{1}{c}{PA}&                \mc{1}{c}{P}&                 \mc{1}{c}{E}&                 
 \mc{1}{c}{I}&                 \mc{1}{c}{E}&                 \mc{1}{c}{C}&                 \mc{1}{c}{P}&                 
 \mc{1}{c}{HV}&                \mc{1}{c}{R}&                 \mc{1}{c}{LON}&               \mc{1}{c}{LAT}&               
 \mc{1}{c}{D$_{\rm Jup}$}&     \mc{1}{c}{ROT}&               \mc{1}{c}{S$_{\rm LON}$}&     \mc{1}{c}{S$_{\rm LAT}$}&     
 \mc{1}{c}{V}&                 \mc{1}{c}{VEF}&               \mc{1}{c}{M}&                 \mc{1}{c}{MEF}\\
\end{tabular*} \end{sidewaystable}                                                     
\clearpage                                                                      
\thispagestyle{empty}
\begin{sidewaystable}                                                                   \
\tiny \hspace{-2cm}
\tabcolsep1.2mm
\begin{tabular*}{23cm} {@{\extracolsep{\fill}}                                  
rrrrrrrrrrrrrrrcrrrrcrrrrrrrrrrrrrrrrrrrrr}                                     
 \mc{1}{c}{No.}&               \mc{1}{c}{IMP. DATE}&         \mc{1}{c}{C}&                 \mc{1}{c}{AR}&                
 \mc{1}{c}{S}&                 \mc{1}{c}{IA}&                \mc{1}{c}{EA}&                \mc{1}{c}{CA}&                
 \mc{1}{c}{IT}&                \mc{1}{c}{ET}&                \mc{1}{c}{E}&                 \mc{1}{c}{E}&                 
 \mc{1}{c}{I}&                 \mc{1}{c}{PA}&                \mc{1}{c}{P}&                 \mc{1}{c}{E}&                 
 \mc{1}{c}{I}&                 \mc{1}{c}{E}&                 \mc{1}{c}{C}&                 \mc{1}{c}{P}&                 
 \mc{1}{c}{HV}&                \mc{1}{c}{R}&                 \mc{1}{c}{LON}&               \mc{1}{c}{LAT}&               
 \mc{1}{c}{D$_{\rm Jup}$}&     \mc{1}{c}{ROT}&               \mc{1}{c}{S$_{\rm LON}$}&     \mc{1}{c}{S$_{\rm LAT}$}&     
 \mc{1}{c}{V}&                 \mc{1}{c}{VEF}&               \mc{1}{c}{M}&                 \mc{1}{c}{MEF}\\
\end{tabular*} \end{sidewaystable}                                                     
\clearpage                                                                      
\thispagestyle{empty}
\begin{sidewaystable}                                                                   \
\tiny \hspace{-2cm}
\tabcolsep1.2mm
\begin{tabular*}{23cm} {@{\extracolsep{\fill}}                                  
rrrrrrrrrrrrrrrcrrrrcrrrrrrrrrrrrrrrrrrrrr}                                     
 \mc{1}{c}{No.}&               \mc{1}{c}{IMP. DATE}&         \mc{1}{c}{C}&                 \mc{1}{c}{AR}&                
 \mc{1}{c}{S}&                 \mc{1}{c}{IA}&                \mc{1}{c}{EA}&                \mc{1}{c}{CA}&                
 \mc{1}{c}{IT}&                \mc{1}{c}{ET}&                \mc{1}{c}{E}&                 \mc{1}{c}{E}&                 
 \mc{1}{c}{I}&                 \mc{1}{c}{PA}&                \mc{1}{c}{P}&                 \mc{1}{c}{E}&                 
 \mc{1}{c}{I}&                 \mc{1}{c}{E}&                 \mc{1}{c}{C}&                 \mc{1}{c}{P}&                 
 \mc{1}{c}{HV}&                \mc{1}{c}{R}&                 \mc{1}{c}{LON}&               \mc{1}{c}{LAT}&               
 \mc{1}{c}{D$_{\rm Jup}$}&     \mc{1}{c}{ROT}&               \mc{1}{c}{S$_{\rm LON}$}&     \mc{1}{c}{S$_{\rm LAT}$}&     
 \mc{1}{c}{V}&                 \mc{1}{c}{VEF}&               \mc{1}{c}{M}&                 \mc{1}{c}{MEF}\\
\end{tabular*} \end{sidewaystable}                                                     
\clearpage                                                                      
\thispagestyle{empty}
\begin{sidewaystable}                                                                   \
\tiny \hspace{-2cm}
\tabcolsep1.2mm
\begin{tabular*}{23cm} {@{\extracolsep{\fill}}                                  
rrrrrrrrrrrrrrrcrrrrcrrrrrrrrrrrrrrrrrrrrr}                                     
 \mc{1}{c}{No.}&               \mc{1}{c}{IMP. DATE}&         \mc{1}{c}{C}&                 \mc{1}{c}{AR}&                
 \mc{1}{c}{S}&                 \mc{1}{c}{IA}&                \mc{1}{c}{EA}&                \mc{1}{c}{CA}&                
 \mc{1}{c}{IT}&                \mc{1}{c}{ET}&                \mc{1}{c}{E}&                 \mc{1}{c}{E}&                 
 \mc{1}{c}{I}&                 \mc{1}{c}{PA}&                \mc{1}{c}{P}&                 \mc{1}{c}{E}&                 
 \mc{1}{c}{I}&                 \mc{1}{c}{E}&                 \mc{1}{c}{C}&                 \mc{1}{c}{P}&                 
 \mc{1}{c}{HV}&                \mc{1}{c}{R}&                 \mc{1}{c}{LON}&               \mc{1}{c}{LAT}&               
 \mc{1}{c}{D$_{\rm Jup}$}&     \mc{1}{c}{ROT}&               \mc{1}{c}{S$_{\rm LON}$}&     \mc{1}{c}{S$_{\rm LAT}$}&     
 \mc{1}{c}{V}&                 \mc{1}{c}{VEF}&               \mc{1}{c}{M}&                 \mc{1}{c}{MEF}\\
\end{tabular*} \end{sidewaystable}                                                     
\clearpage                                                                      
\thispagestyle{empty}
\begin{sidewaystable}                                                                   \
\tiny \hspace{-2cm}
\tabcolsep1.2mm
\begin{tabular*}{23cm} {@{\extracolsep{\fill}}                                  
rrrrrrrrrrrrrrrcrrrrcrrrrrrrrrrrrrrrrrrrrr}                                     
 \mc{1}{c}{No.}&               \mc{1}{c}{IMP. DATE}&         \mc{1}{c}{C}&                 \mc{1}{c}{AR}&                
 \mc{1}{c}{S}&                 \mc{1}{c}{IA}&                \mc{1}{c}{EA}&                \mc{1}{c}{CA}&                
 \mc{1}{c}{IT}&                \mc{1}{c}{ET}&                \mc{1}{c}{E}&                 \mc{1}{c}{E}&                 
 \mc{1}{c}{I}&                 \mc{1}{c}{PA}&                \mc{1}{c}{P}&                 \mc{1}{c}{E}&                 
 \mc{1}{c}{I}&                 \mc{1}{c}{E}&                 \mc{1}{c}{C}&                 \mc{1}{c}{P}&                 
 \mc{1}{c}{HV}&                \mc{1}{c}{R}&                 \mc{1}{c}{LON}&               \mc{1}{c}{LAT}&               
 \mc{1}{c}{D$_{\rm Jup}$}&     \mc{1}{c}{ROT}&               \mc{1}{c}{S$_{\rm LON}$}&     \mc{1}{c}{S$_{\rm LAT}$}&     
 \mc{1}{c}{V}&                 \mc{1}{c}{VEF}&               \mc{1}{c}{M}&                 \mc{1}{c}{MEF}\\
\end{tabular*} \end{sidewaystable}                                                     
\clearpage                                                                      
\thispagestyle{empty}
\begin{sidewaystable}                                                                   \
\tiny \hspace{-2cm}
\tabcolsep1.2mm
\begin{tabular*}{23cm} {@{\extracolsep{\fill}}                                  
rrrrrrrrrrrrrrrcrrrrcrrrrrrrrrrrrrrrrrrrrr}                                     
 \mc{1}{c}{No.}&               \mc{1}{c}{IMP. DATE}&         \mc{1}{c}{C}&                 \mc{1}{c}{AR}&                
 \mc{1}{c}{S}&                 \mc{1}{c}{IA}&                \mc{1}{c}{EA}&                \mc{1}{c}{CA}&                
 \mc{1}{c}{IT}&                \mc{1}{c}{ET}&                \mc{1}{c}{E}&                 \mc{1}{c}{E}&                 
 \mc{1}{c}{I}&                 \mc{1}{c}{PA}&                \mc{1}{c}{P}&                 \mc{1}{c}{E}&                 
 \mc{1}{c}{I}&                 \mc{1}{c}{E}&                 \mc{1}{c}{C}&                 \mc{1}{c}{P}&                 
 \mc{1}{c}{HV}&                \mc{1}{c}{R}&                 \mc{1}{c}{LON}&               \mc{1}{c}{LAT}&               
 \mc{1}{c}{D$_{\rm Jup}$}&     \mc{1}{c}{ROT}&               \mc{1}{c}{S$_{\rm LON}$}&     \mc{1}{c}{S$_{\rm LAT}$}&     
 \mc{1}{c}{V}&                 \mc{1}{c}{VEF}&               \mc{1}{c}{M}&                 \mc{1}{c}{MEF}\\
\end{tabular*} \end{sidewaystable}                                                     
}


\clearpage

\section*{Figures}

\vskip -2cm

\nocite{landgraf2000b}

\end{document}